\documentclass[a4paper,fleqn]{cas-sc}

\usepackage[authoryear]{natbib}
\usepackage{pdfpages}
\usepackage{subfigure}
\usepackage{amssymb}
\usepackage{amsmath}
\usepackage{textcomp}
\usepackage{bbding}
\usepackage{url}
\usepackage{lineno}
\usepackage{booktabs}
\usepackage{threeparttable}
\usepackage{setspace}

\def\tsc#1{\csdef{#1}{\textsc{\lowercase{#1}}\xspace}}
\tsc{WGM}
\tsc{QE}
\tsc{EP}
\tsc{PMS}
\tsc{BEC}
\tsc{DE}

\begin{document}
\let\WriteBookmarks\relax
\def\floatpagepagefraction{1}
\def\textpagefraction{.001}
\let\printorcid\relax

\shorttitle{Enhancing Terrestrial NPP Estimation with EXP-CASA}

\shortauthors{G. Chen, et~al.}

\title [mode = title]{Enhancing Terrestrial Net Primary Productivity Estimation with EXP-CASA: A Novel Light Use Efficiency Model Approach}

\author[inst1]{Guanzhou Chen}
\credit{Conceptualization, Resources, Writing - original draft}
\fnmark[1]
\author[inst1,inst2]{Kaiqi Zhang}
\credit{Methodology, Data Curation, Visualization, Writing - review \& editing}
\fnmark[1]
\author[inst1]{Xiaodong Zhang}
\credit{Supervision, Project administration, Funding acquisition}
\ead{zxdlmars@whu.edu.cn}
\cormark[1]
\author[inst3]{Hong Xie}
\credit{Writing - review \& editing, Visualization}
\author[inst4]{Haobo Yang}
\credit{Formal analysis, Validation, Writing - review \& editing}
\author[inst1]{Xiaoliang Tan}
\credit{Investigation, Validation, Writing - review \& editing}
\author[inst1]{Tong Wang}
\credit{Validation, Writing - review \& editing, Formal analysis}
\author[inst2]{Yule Ma}
\credit{Formal analysis, Writing - review \& editing}
\author[inst3]{Qing Wang}
\credit{Investigation, Writing - review \& editing}
\author[inst5]{Jinzhou Cao}
\credit{Investigation, Writing - review \& editing}
\author[inst2]{Weihong Cui}
\credit{Supervision, Writing - review \& editing}

\affiliation[inst1]{organization={State Key Laboratory of Information Engineering in Surveying, Mapping and Remote Sensing, Wuhan University},
            addressline={No.129, Luoyu Road}, 
            city={Wuhan},
            postcode={420079}, 
            state={Hubei},
            country={China}}
\affiliation[inst2]{organization={School of Remote Sensing and Information Engineering, Wuhan University},
            addressline={No.129, Luoyu Road}, 
            city={Wuhan},
            postcode={420079}, 
            state={Hubei},
            country={China}}
\affiliation[inst3]{organization={School of Geosciences, Yangtze University},
            addressline={111 University Road, Caidian Street, Caidian district},
            city={Wuhan},
            postcode={430100},
            state={Hubei},
            country={China}}
\affiliation[inst4]{organization={PowerChina Kunming Engineering Corporation Limited},
            addressline={No.115, East Renmin Road},
            city={Kunming},
            postcode={650051},
            state={Yunnan},
            country={China}}
\affiliation[inst5]{organization={Shenzhen Technology University},
            addressline={College of Big Data and Internet}, 
            city={Shenzhen},
            postcode={518118}, 
            state={Guangdong},
            country={China}}

\cortext[cor1]{Corresponding author}

\fntext[fn1]{G. Chen and K. Zhang contributed equally to this work.}

\begin{abstract}
The Light Use Efficiency (LUE) model, epitomized by the Carnegie-Ames-Stanford Approach (CASA) model, is extensively applied in the quantitative estimation and analysis of vegetation Net Primary Productivity (NPP). However, the classic CASA model is marked by significant complexity: the estimation of environmental stress parameters, in particular, necessitates multi-source observation data, adding to the complexity and uncertainty of the model's operation. Additionally, the saturation effect of the Normalized Difference Vegetation Index (NDVI), a key variable in the CASA model, weakened the accuracy of CASA's NPP predictions in densely vegetated areas. To address these limitations, this study introduces the Exponential-CASA (EXP-CASA) model. The EXP-CASA model effectively improves the CASA model by using novel functions for estimating the fraction of absorbed photosynthetically active radiation (FPAR) and environmental stress, by utilizing long-term observational data from FLUXNET and MODIS surface reflectance data. In a comparative analysis of NPP estimation accuracy among four different NPP products, EXP-CASA ($R^2 = 0.68, RMSE= 1.1 gC\cdot m^{-2} \cdot d^{-1}$) outperforms others, followed by GLASS-NPP ($R^2 = 0.61, RMSE= 1.2 gC\cdot m^{-2} \cdot d^{-1}$), and lastly MODIS-NPP ($R^2 = 0.46, RMSE= 1.77 gC\cdot m^{-2} \cdot d^{-1}$) and classic CASA ($R^2 = 0.25, RMSE= 1.76 gC\cdot m^{-2} \cdot d^{-1}$). Additionally, this research assesses the EXP-CASA model's adaptability to various vegetation indices, evaluates the sensitivity and stability of its parameters over time, and compares its accuracy against other leading NPP estimation products across different seasons, latitudinal zones, ecological types, and temporal sequences. The findings reveal that the EXP-CASA model exhibits strong adaptability to diverse vegetation indices and stability of model parameters over time series. Importantly, EXP-CASA displays superior sensitivity to NPP anomalies at flux sites and more accurately simulates short-term NPP fluctuations and captures periodic trends. By introducing a novel estimation approach that optimizes model construction, the EXP-CASA model remarkably improves the accuracy of NPP estimations and paves the way for global-scale, consistent, and continuous assessment of vegetation NPP and presents an effective approach for evaluating the saturation effect of vegetation indices and the influence of category independence on the NPP estimation.
\end{abstract}



\maketitle

\onehalfspacing

\section{Introduction}
Over the last few decades, a notable increase in greenhouse gases, especially carbon dioxide, has been observed in the Earth's atmosphere, leading to global warming \citep{zhang2021carbon,liu2022carbon}. In response, significant research emphasis has been placed on ecosystems' capacity for carbon absorption and sequestration \citep{mu2013grassland,mu2013assessing,liu2022detection}. 
Terrestrial vegetation plays an instrumental role in the carbon cycling process, acting as both a vital carbon sink and a barometer for ecosystem health.\citep{piao2005changes,zhu2005npp,zhu2007npp,seiler2022terrestrial, cai2023detecting,zhu2024remote}. 
Central to understanding this biological process is the metric of Net Primary Productivity (NPP), which quantifies the net carbon sequestration by vegetation. NPP is determined by measuring the balance of CO$_2$ absorption from the atmosphere through photosynthesis against its release via autotrophic respiration. This metric is essential for measuring the efficiency of carbon capture by terrestrial vegetation and provides insights into the dynamics of the global carbon cycle
\citep{franklin2016global,xu2019increasing,ning2021land,piao2022carbon}.

In recent years, a diverse range of methodologies including climate models, statistical methods, machine learning methods, and light use efficiency (LUE) models have been employed to enhance the accuracy of NPP estimations from spatial and temporal perspectives \citep{pruavualie2023machine}. However, each of these methodologies comes with inherent limitations that compromise the accuracy and applicability of NPP estimations \citep{bao2016modeling}. Climate models involve establishing regression models based on the statistical relationships between climatic factors such as temperature, precipitation and the measured biomass \citep{lieth1973primary, lieth1975modeling, uchijima1985agroclimatic}. Its estimation accuracy is significantly affected by the low spatial and temporal resolution of the climatic factors \citep{deyong2009does,yu2009modelling,bao2016modeling}. Statistical methods, which are based on vegetation indices, are straightforward but suffer from poor estimation accuracy and saturation effects in high vegetation density areas \citep{gao2023evaluating,wang2024progress}.  Meanwhile, machine learning approaches, despite their strength in fitting complex patterns, are prone to be difficult to balance between the interpretability of model parameters and the accuracy of model estimations, posing challenges in ecological studies \citep{reichstein2019deep,pruavualie2023machine,wang2024progress}. 

Grounded on theoretical foundations and empirical studies, 
the light use efficiency (LUE) models have demonstrated their explicability and potential in NPP estimation, particularly benefiting from advancements in remote sensing technology \citep{bao2016modeling,chen2020spatial,qi2023net,wang2024progress}. LUE models, such as the Carnegie-Ames-Stanford-Approach (CASA) \citep{potter1993terrestrial,piao2005changes,wang2020spatio,yin2022contributions}, the Global Production Efficiency Model (GLO-PEM) \citep{prince1991model}, the 3-PG model \citep{landsberg1997generalised}, and the EC-LUE model \citep{sims2005midday}, offer improved accuracy and interpretability compared to traditional statistical methods, but still rely heavily on multi-source meteorological data \citep{running1993generalization}. This research specifically seeks to address three critical challenges in current LUE models for NPP estimation: (1) the saturation effect of vegetation indices  \citep{gao2023evaluating}, (2) the low spatial resolution and accuracy in environmental stress estimations based on meteorological data \citep{potter1993terrestrial,zhu2006simulation,zhu2007npp}, and (3) the complexity in parameterizing and calibrating the estimation model due to its characteristic as a multiplicative model involving several factors \citep{piao2005changes,yu2009modelling,bao2016modeling}. This study presents a novel model, the Exponential-CASA (EXP-CASA) model, derived from the analysis of the classical CASA model's characteristics in logarithmic space. Meanwhile, we evaluates the stability and effectiveness of the improved model, and conducts accuracy comparison with major NPP products in site level based on the global flux network (FLUXNET) data.

Our research primarily makes three contributions. (1) This study presents the EXP-CASA model, which incorporates refined methods for estimating the fraction of absorbed photosynthetically active radiation (FPAR), water stress ($W_s$), and temperature stress ($T_s$). These enhancements have increased the accuracy of NPP estimations. (2) By separating the model parameters from its inputs and refining the CASA model through conducting multiple least squares regression in logarithmic space, we not only further simplified the framework of the estimation model and reduced its computational complexity but also effectively ensured the physical significance and interpretability of each component within the light use efficiency model. (3) Utilizing the EXP-CASA methodology, this study conducted a deeper analysis of the impacts of global consistent factors on the NPP estimation and quantitatively assessed the vegetation index's saturation effect. This enabled an effective evaluation of EXP-CASA's advantages and limitations.

The rest of this paper is organized as follows. Section \ref{Related works} provides related works on the LUE models for estimating NPP. Section \ref{Materials and methods} comprehensively describes the data and the improved EXP-CASA model proposed in this study. Section \ref{Results} presents the experimental results. Section \ref{Discussions} analyzes the experimental results. Section \ref{Conclusions} summarizes the contributions of this study to improving the methods for estimating net primary productivity and proposes directions for future research.

\section{Related works}
\label{Related works}
\subsection{Light use efficiency (LUE) model}
The LUE model simplifies and abstracts the photosynthesis of vegetation: the strength of vegetation photosynthesis is related to the amount of effective solar radiation absorbed by leaves, and environmental factors will limit the potential light use efficiency and thus affect the photosynthesis \citep{potter1993terrestrial,field1995global,zhu2007npp}. Taking the typical LUE model CASA as an example:
\begin{equation}
    \label{eq:CASA}
    \begin{aligned}
    NPP &= APAR \times LUE \\
    APAR &= Rad \times FPAR \times 0.5 \\
    LUE &= LUE_{max} \times T_{s} \times W_{s} 
    \end{aligned}
\end{equation}
where $APAR$ represents absorbed photosynthetic active radiation (unit: $MJ\cdot m^{-2}$); $FPAR$ represents fraction of absorbed photosynthetic active radiation; the factor 0.5 takes into account the reality that roughly half of the incoming solar radiation \citep{potter1993terrestrial,field1995global,zhu2007npp}; $Rad$ represents total solar radiation (unit: $MJ\cdot m^{-2}$); $LUE_{max}$ represents the maximum light use efficiency ($gC\cdot MJ^{-1}$); $T_{s},W_{s}$ represent temperature and water stress coefficients on actual light use efficiency, respectively. 

Most LUE models estimate FPAR by using either the vegetation index or a linear transformation of the vegetation index as its direct proxy \citep{landsberg1997generalised,sims2005midday,zhu2007npp,su2022optimization,chang2022effects}. Recent studies on the estimation of FPAR start from the perspective of vegetation indices, with specific details provided in Section \ref{fpar}. Meanwhile, actual light use efficiency is composed of two parts: one part is the maximum light use efficiency (see Section \ref{lue_optimization}), which is a metric related to the vegetation category, and the other part is photosynthetic suppression caused by environmental factors. 

Water stress ($W_s$) is used to reflect the influence of the effective moisture conditions available to vegetation on light use efficiency. In the theoretical estimation of water stress, it is necessary to comprehensively consider the ecosystem evapotranspiration formed by vegetation transpiration and soil evaporation \citep{potter1993terrestrial,field1995global,zhu2007npp}. 
Taking the classic CASA model as an example, temperature, precipitation, and soil texture data are used to estimate the soil moisture content and then calculate the evapotranspiration to estimate the degree of moisture stress \citep{potter1993terrestrial,field1995global}. However, this estimation method involves complex parameters, and its spatial resolution is usually low, which may affect the precision level of primary productivity estimation \citep{potter1993terrestrial,yu2009modelling,bao2016modeling,su2022optimization}.

Temperature stress ($T_s$) is used to describe the fact that LUE of vegetation inevitably decreases under conditions that deviate from the optimal growth temperature \citep{potter1993terrestrial,field1995global,stocker2019drought,qiao2024impact}. 
For the estimation of temperature stress, a piecewise function was proposed. This function is based on the actual environmental temperature and the optimal growth temperature of vegetation to describe the stress \citep{potter1993terrestrial,field1995global}. This function effectively captures how vegetation photosynthesis is constrained by both extreme temperatures and the optimal temperature range for vegetation growth \citep{lieth1973primary,lieth1975modeling,uchijima1985agroclimatic,piao2005changes}. It is noted that estimating the optimal growth temperatures for vegetation species via remote sensing techniques typically hinges on comprehensive analysis of observational datasets and specific vegetation indices \citep{zhu2006simulation,piao2005changes}. Nevertheless, the potential temporal delay between the vegetation's ideal growth temperatures and the corresponding indicators from vegetation indices introduces an additional layer of complexity \citep{li2021ecostress,sun2024high}. This challenge complicates the process of enacting swift adjustments in response to the dynamic shifts of primary productivity within terrestrial vegetation ecosystems \citep{li2023considering,sun2024high}.

\subsection{Application of nonlinear vegetation indices to primary productivity estimation}
\label{fpar}
While many large-scale vegetation primary productivity estimation models consider vegetation indices to be a good proxy for the fraction of absorbed photosynthetically active radiation (FPAR) \citep{potter1993terrestrial,field1995global,sims2005midday,zhu2007npp,su2022optimization,chang2022effects}, the saturation effect of classical vegetation indices like NDVI is hard to ignore \citep{camps2021unified,gao2023evaluating}. This effect is especially prominent in areas with high vegetation canopy cover, where a simple linear proxy can lead to significant estimation errors \citep{gao2023evaluating,qi2023net}. Non-linear vegetation indices have shown good performance across various ecological categories, in analyses related to Solar-induced Chlorophyll Fluorescence (SIF) \citep{camps2021unified,zeng2022optical,gao2023evaluating}. This offers new perspectives for estimating vegetation productivity and addressing saturation effects of vegetation indices. Given that kNDVI and NIR$_v$ have been demonstrated to effectively mitigate saturation effects \citep{camps2021unified,gao2023evaluating}, this study incorporates them as the chosen vegetation indices within the model.

NIR$_v$ is a nonlinear vegetation index proposed based on NDVI, which is obtained by calculating the product of NDVI and the corresponding near-infrared band reflectance \citep{badgley2017canopy,dechant2022nirvp}. The calculation method is as follows:
\begin{equation}
    \label{eq:NIR_v}
    NIR_v = NDVI \times \rho_{NIR} = \frac{\rho_{NIR} - \rho_{RED}}{\rho_{NIR} + \rho_{RED}} \times \rho_{NIR}
\end{equation}

where $\rho_{NIR}$ and $\rho_{RED}$ represent the surface reflectance of the near-infrared and red bands, respectively. NIR$_v$ has been proven to alleviate the saturation effect of NDVI to a certain extent, but its performance is poor in areas with high vegetation coverage \citep{camps2021unified,gao2023evaluating}. Simultaneously, the derivative of NIR$_v$ to NDVI represents the surface reflectance in the near-infrared band. Its sensitivity increases linearly with $\rho_{NIR}$, however, this derivative's capability to address the saturation issue inherent in nonlinear functions is generally limited \citep{camps2021unified}.
 
In the course of studying the relationship between nonlinear vegetation indices and vegetation primary productivity, the kNDVI was proposed, and it has been found to be another good proxy for primary productivity \citep{camps2021unified}. The calculation method is as follows:
\begin{equation}
    \label{eq:kNDVI}
    kNDVI = \tanh\left ( \left ( \frac{\rho_{NIR}-\rho_{RED}}{2\sigma } \right )^{2} \right ) 
\end{equation}
where $\rho_{NIR}$ and $\rho_{RED}$ respectively represent the surface reflectance of the near-infrared and red bands; $\sigma$ represents the sensitivity of the index to regional vegetation density. On this basis, two forms of the equation are derived: one uses the global approximation value of 0.15 obtained in experiments as a fixed $\sigma$; the other uses the spatially heterogeneous $0.5\cdot(\rho_{NIR} + \rho_{RED})$ as the $\sigma$ value \citep{camps2021unified}, thereby deriving the simplified equation as follows:
\begin{equation}
    \label{eq:kNDVI_Simplified}
    kNDVI = \tanh\left( NDVI ^{2}\right) 
\end{equation}

Currently, nonlinear vegetation indices are widely used to estimate vegetation primary productivity. Notably, the kNDVI demonstrates a strong correlation with SIF in regions of 
high dense vegetation canopy, underscoring its potential for estimating primary productivity \citep{camps2021unified}. Some research has used kNDVI as a proxy for FPAR to optimize the CASA model for estimating the net primary productivity of vegetation in China, proving that the accuracy of net primary productivity simulations by the CASA model was improved after optimization with kNDVI \citep{qi2023net}. However, merely altering the FPAR vegetation index proxy does not adequately optimize the CASA model, and its potential for global application remains unverified.

\subsection{Optimization of actual light use efficiency}
\label{lue_optimization}
The majority of recent studies aimed at enhancing actual light use efficiency concentrate on moisture stress and maximum light use efficiency (LUE$_{max}$). The land surface water index (LSWI)-based stress estimation methodology has emerged as the predominant approach for refining water stress estimations, and its application value  has been confirmed through related studies \citep{bao2016modeling,su2022optimization}.

The shortwave infrared (SWIR) spectral band is sensitive to vegetation water content and soil moisture, and the LSWI is derived by combining $\rho_{NIR}$ and $\rho_{SWIR}$ \citep{xiao2002characterization,xiao2004modeling,xiao2004satellite,wu2022improved}. LSWI is based on surface reflectance data, which has a high spatial and temporal resolution and is easy to obtain. As soil moisture content increases, leaf reflectance of $\rho_{SWIR}$ decreases, leading to an increase in LSWI values \citep{xiao2002characterization,xiao2004satellite}. LSWI is calculated as follows:
\begin{equation}
    LSWI = \frac{\rho_{NIR} - \rho_{SWIR}}{\rho_{NIR} + \rho_{SWIR}}
\end{equation}
where $\rho_{NIR}$ and $\rho_{SWIR}$ represent the near-infrared (NIR) band and short wave infrared (SWIR) band, respectively. LSWI-based water stress was calculated as follows \citep{bao2016modeling,su2022optimization}:
\begin{equation}
    W_{s} = \frac{1 + LSWI}{1 + LSWI_{max}}
\end{equation}
where $LSWI_{max}$ represents the maximum LSWI during the period of vegetation growth.

LSWI can be calculated from surface reflectance, which has a higher spatial and temporal resolution and is less dependent on hyperparameters than evapotranspiration estimation using precipitation data and soil texture data \citep{bao2016modeling,su2022optimization}. It is noted that LSWI-based water stress improvement still relies on long time-series observations of vegetation growth period data and is difficult to reflect the inhibitory effect of vegetation photosynthesis under extreme water conditions \citep{xiao2005modeling,xiao2005satellite,bao2016modeling}.

Meanwhile, the estimation of LUE$_{max}$ is primarily performed through the observation and analysis of biomass data across specific categories, with direct estimates based on statistical biomass data \citep{running2000global,zhu2006simulation,liu2019global,xiao2022estimation}. However, the effectiveness of LUE$_{max}$ estimates can vary depending on the models used, and factors such as geographic area, spatial resolution, and vegetation cover categories play a significant role in influencing the outcomes. Moreover, discernible spatial variations within these studies' datasets, highlighted by data specific to regions such as China \citep{zhu2006simulation} or North America \citep{running2000global}, indicate that spatial discrepancies in LUE$_{max}$ have not been fully eliminated. Concurrently, the use of different vegetation ecological classification systems across studies introduces further complexity \citep{running2000global,zhu2006simulation,bao2016modeling,liu2019global,su2022optimization}. The lack of clarity poses substantial challenges in developing consistent and accurate models of global light use efficiency.

\section{Materials and methods}
\label{Materials and methods}
\subsection{Model input data}
\subsubsection{Surface  reflectance and meteorological data}
The surface reflectance data for the model consists of two products: the MCD43A4.061 surface reflectance product and the MCD43A2.061 quality control product \citep{Schaaf2021-ma}. The surface reflectance data has been adjusted using the Bidirectional Reflectance Distribution Function (BRDF-Adjusted) to eliminate the impact of directional reflection from the ground surface. The quality control product provides additional auxiliary information for the MCD43A4 product, thus eliminating low-quality BRDF estimation data. The surface reflectance data, following aggregation, are resampled to a 500-meter resolution and subsequently averaged into an 8-day mean via mean aggregation. For each site, surface reflectance data are derived from the median within a 500-meter radius buffer surrounding the site locations, maintaining a temporal resolution of 8 days.

The meteorological inputs for the model include solar radiation and surface temperature, which comes from the ERA5-Land Daily Aggregated dataset within the Copernicus Climate Change Service (C3S) and Climate Data Store (CDS) \citep{munoz2019era5}. This dataset integrates simulated data and global observation data, with a spatial and temporal resolution of 0.1\textdegree and 1 day, respectively. The preprocessing methods for meteorological data are the same as those for surface reflectance data.

\subsubsection{FLUXNET data: eddy covariance flux observations}

\begin{figure}[ht]
    \centering
    \includegraphics[width=\textwidth]{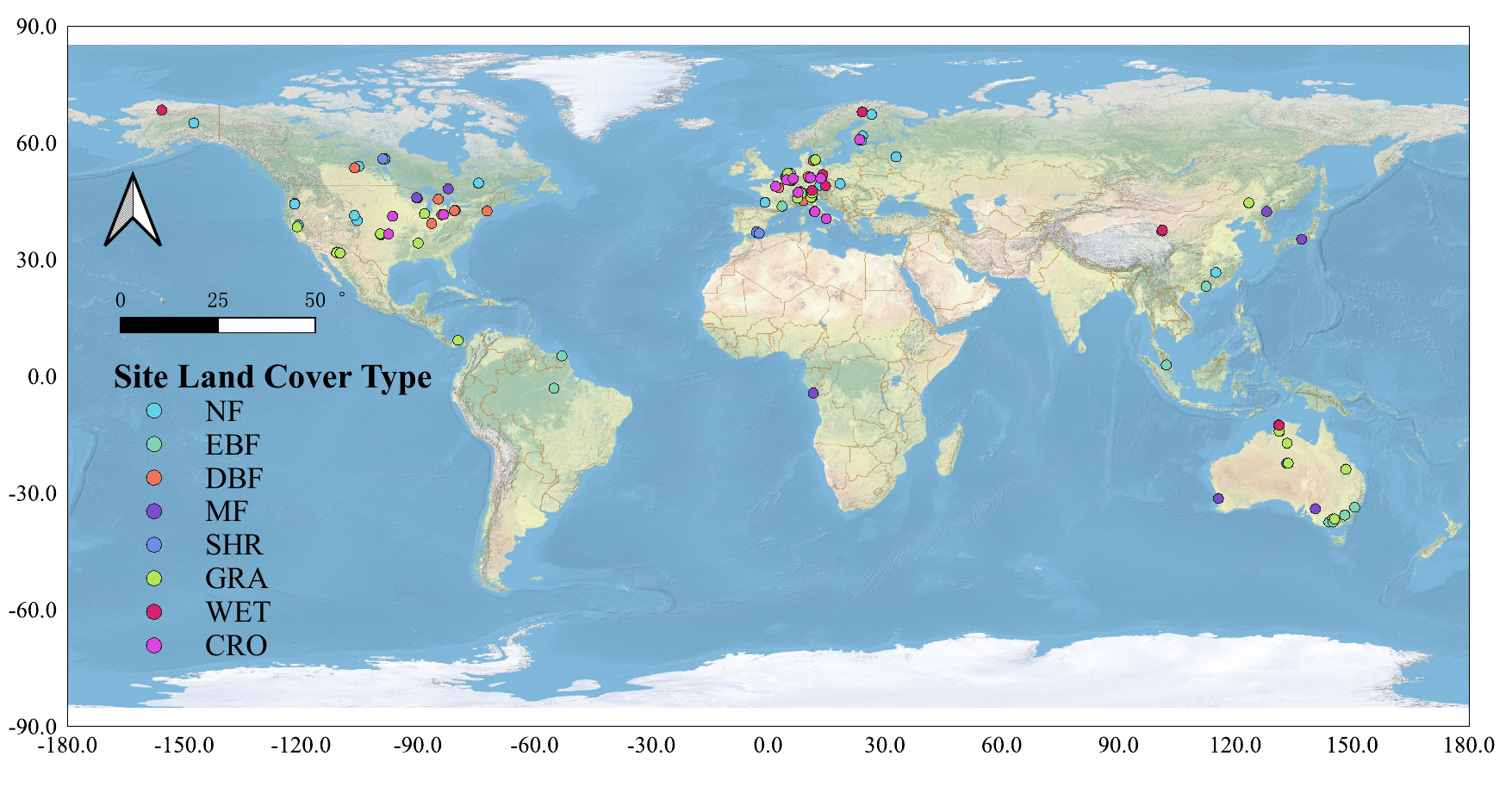}
    \caption{Geographical distribution of FLUXNET sites utilized in this study. Site classifications are based on land cover categories recorded at the sites, encompassing eight principal categories: Needleleaf Forest (NF), Evergreen Broadleaf Forest (EBF), Deciduous Broadleaf Forest (DBF), Mixed Forest (MF), Shrubland (SHR), Grassland (GRA), Permanent Wetland (WET), and Cropland (CRO).}
    \label{fig:SiteDistribution}
\end{figure}

In this study, eddy covariance flux data were utilized to calibrate the model and assess the accuracy of the NPP estimations generated by the refined model. Data for this analysis was sourced from the open sites within the FLUXNET network \citep{pastorello_fluxnet2015_2020, mallick2024net}, adhering to strict quality control measures: excluding sites with fewer than three years of continuous observation or with less than 65\% high-quality data, as well as eliminating any sites lacking observational records. Consequently, the final dataset comprised 121 sites, spanning eight primary vegetation categories to this research (detailed in Appendix \ref{appendix_A}). These included 27 needleleaf forest (NF), 10 evergreen broadleaf forest (EBF), 16 deciduous broadleaf forest (DBF), 15 mixed forest (MF), 6 shrubland (SH), 25 grassland (GRA), 7 permanent wetland (WET), and 15 cropland (CRO) sites, with their geographical distribution depicted in Figure \ref{fig:SiteDistribution}.

FLUXNET provides net ecosystem exchange (NEE) measurements at half-hour intervals, aggregated across daily and monthly periods, while gross primary productivity (GPP) is calculated via DAYTIME \citep{lasslop2010separation} and NIGHTTIME \citep{reichstein2005separation} algorithms. Within this framework, the study employs the average GPP values from these algorithms as the observed GPP at the sites. From there, site-specific observed NPP data were derived based on established GPP to NPP relations \citep{waring1998net,landsberg2020assessment,chang2022effects}, underpinning this research's analytical foundation.

\subsection{EXP-CASA: exponential-CASA model}
\begin{figure}[ht]
    \centering
    \includegraphics[width=\textwidth]{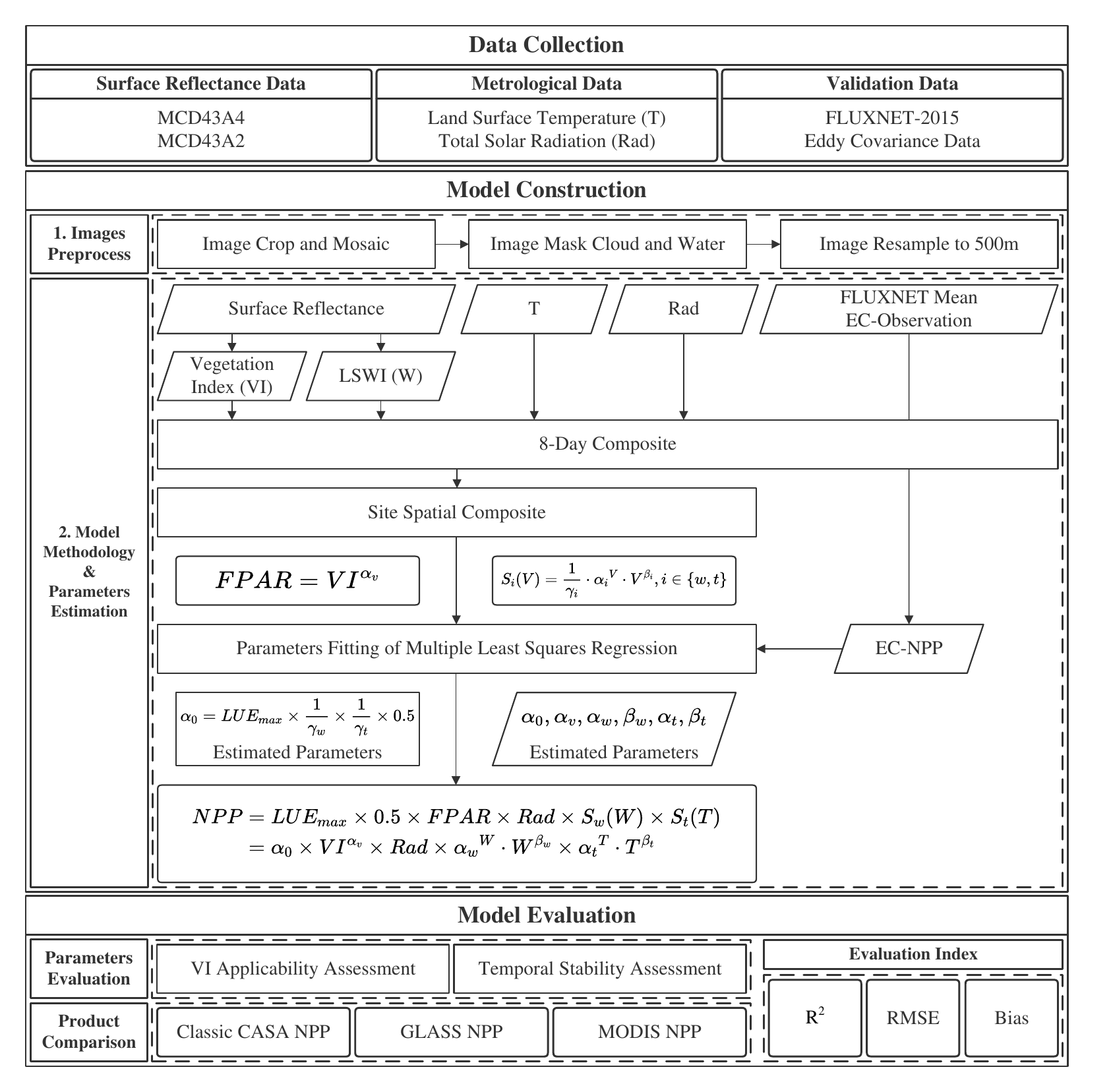}
    \caption{Technical Workflow of this Study. This diagram delineates the three major stages involved in the research: Data Collection, Model Construction, and Model Evaluation. It provides a concise overview of the methodology, from the initial aggregation of datasets to the development of the analytical model, and culminating in a thorough evaluation of the model's effectiveness.}
    \label{fig:Methods_frame}
\end{figure}

This study's methodology comprises three main phases, as depicted in Figure \ref{fig:Methods_frame}. Initially, the data acquisition phase involves collecting MODIS surface reflectance data, meteorological data, FLUXNET data, and other NPP products, with further details available in Appendix \ref{appendix_A}. Subsequently, the second phase focuses on refining the classical CASA model by introducing novel methods for estimating FPAR and environmental stress. The final phase evaluates the model's effectiveness, assessing both the stability of its parameters and the accuracy of its NPP estimations.

In our research, based on the classic CASA model, we improved the estimation methods for FPAR and LUE and used flux observation data for model calibration, proposing a novel model for estimating net primary productivity. This model reconstructs the calculation methods for FPAR and actual LUE. Since its parameters are linearly fitted in logarithmic space, we refer to this model as the Exponential-CASA model (EXP-CASA).

\subsubsection{Enhancements in FPAR estimation}
Generally, there is a nearly linear correlation between FPAR and the vegetation index \citep{zeng2022optical}, and many studies use NDVI as its direct proxy, such as in the CASA model, 3PG model, EC-LUE model \citep{sims2005midday,zhu2007npp,su2022optimization,chang2022effects}. However, the saturation effect of vegetation indices is widespread, and it is more pronounced in areas with high vegetation canopy density - using NDVI as a direct proxy for FPAR will obviously produce larger bias in areas of high vegetation density \citep{gao2023evaluating}. Although a number of nonlinear vegetation indices, such as NIR$_v$ and kNDVI, have been proven to handle the nonlinear saturation effect issue well \citep{camps2021unified, gao2023evaluating}, the quantitative relationship between them and FPAR as well as the extent of suppression of the saturation effect remain difficult to assess. Therefore, this paper proposes a FPAR estimation method based on the form of a power function for the vegetation index, as seen in Equation \ref{eq:FPAR}:
\begin{equation}
FPAR = VI ^ {\alpha_{v}}
\label{eq:FPAR}
\end{equation}
where VI represents vegetation index; $\alpha_{v}$ represents vegetation saturation effect control factor in the form of power function. 

In terrestrial ecosystems, the vegetation index is distributed between 0 and 1. Classic vegetation indices show a good linear correlation with biomass in areas of low to medium vegetation density, whereas in areas of high vegetation density, changes in biomass often outpace changes in the vegetation index, displaying a nonlinear relationship \citep{camps2021unified,zeng2022optical,gao2023evaluating}. Through this form of a power function, this paper conducts a nonlinear mapping based on the vegetation index with only one parameter $\alpha_v$. Although it sacrifices some fitting accuracy in areas of low vegetation density, it can increase the sensitivity of the vegetation index to biomass, thereby effectively improving the accuracy of NPP estimation.

\subsubsection{Improvements in stress assessment}
The stress factor is an indicator used to describe the extent to which photosynthesis in vegetation is inhibited under suboptimal temperature and moisture conditions. When the environment is at the vegetation's optimal growth conditions, the theoretical value of the stress factor is 1, at which point the environment's impact on vegetation photosynthesis is minimized. Conversely, as the environment deviates from the optimal growth conditions for vegetation, environmental factors increasingly impact photosynthesis, manifested as a continuous decrease in the value of the stress factor. In extreme cases, the minimum value of the stress factor can drop to 0.

In this study, in order to simplify the estimation form of stress and effectively fit the effects of different moisture and temperature environmenal conditions on photosynthesis, we proposed a stress estimation method that combines the forms of an exponential function and a power function. The equation is as follows:
\begin{equation}
\label{stress}
S(V) = \frac{1}{\gamma}\cdot{\alpha^{V}\cdot V^{\beta}}
\end{equation}
where $S$ represents stress; $V$ represents normalized temperature or LSWI; $\alpha,\beta$ represent photosynthesis inhibitors; $\gamma$ represents the proportionality adjustment factor, which is used to control stress so that its value range between 0 and 1. 

\begin{figure}[ht]
    \centering
    \includegraphics[width = 0.5\textwidth]{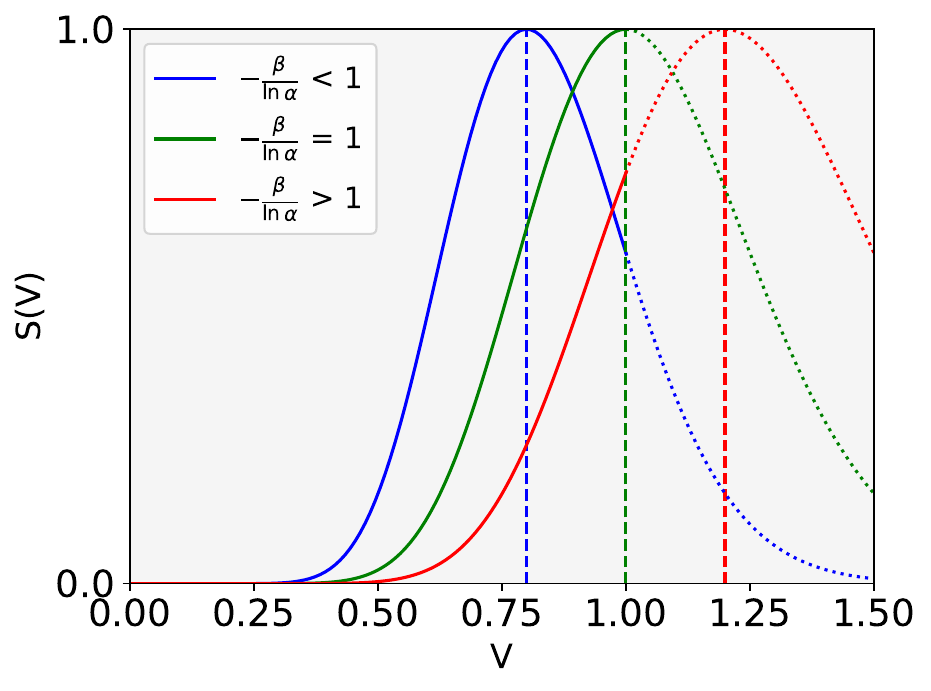}
    \caption{An illustration of stress estimation function in EXP-CASA. This figure presents a function image generated under the assumptions of specific $\alpha$ and $\beta$ values, normalized by gamma. For regions where V exceeds 1, a dotted pattern is employed to depict these sections, whereas the $-\frac{\beta}{\ln \alpha}$ value corresponding to the maximum point is delineated with a dashed line.}
    \label{fig:curveFunction}
\end{figure}

From a mathematical standpoint, the function $S(V)$ is characterized as unimodal: it exhibits minimal values as $V$ nears zero, with the value of $S(V)$ progressively ascending as $V$ increases, subsequent to which it peaks and subsequently declines. By deriving $S(V)$, it is determined that the maximum value is achieved at $-\frac{\beta}{\ln \alpha}$, and by calculating this maximum value, the corresponding $\gamma$ can be determined as follows: 

\begin{equation}
\label{gamma}
\gamma = \alpha^{\frac{\beta}{\ln \alpha}}\cdot (-\frac{\beta}{\ln \alpha})^{-\beta}
\end{equation}
where $\alpha$, $\beta$ hold the same significance as delineated in Equation \ref{stress}. The function curves obtained by simulating environmental conditions at different peak positions are shown in Figure \ref{fig:curveFunction}. Given that the environmental factor $V$ is normalized to ensure the curves' integrity, sections where $V$ exceeds 1 are depicted in a dotted format, while the $-\frac{\beta}{\ln \alpha}$ corresponding to the peak is marked with a dashed line. This curve accurately reflects the strong inhibitory effect at low levels of moisture and temperature conditions and the weak inhibition at high levels, reaching a peak under optimal environmental conditions.

The improved stress curve reflects the process of stress changing with environmental factors: if $-\frac{\beta}{\ln \alpha}$ is less than 1, at lower temperatures or moisture conditions, the photosynthesis of vegetation will be restricted to a very low level; when the environmental conditions reach a certain level, the photosynthetic efficiency of the vegetation will increase non-linearly; upon reaching the optimal growth temperature and moisture conditions for the vegetation, the stress will also peak; when it exceeds $-\frac{\beta}{\ln \alpha}$, the photosynthesis of the vegetation will be inhibited, showing a certain degree of decline but not dropping to a very low level. If $-\frac{\beta}{\ln \alpha}$ is close to 1, the stress curve will show a monotonically increasing state and peak when environmental factors reach to 1. Should $-\frac{\beta}{\ln \alpha}$ exceed 1, the stress curve will exhibit a monotonically increasing trend without reaching a peak, indicating a significant stress impact on vegetation growth. In this study, we use subscripts to denote water stress and temperature stress, which are 
both calculated by Equation \ref{stress}, along with their corresponding parameters ($S_w$ represents water stress, $S_t$ represents temperature stress, etc.).

\subsubsection{Least squares parameter fitting based on flux observation data}

By integrating the above methods, this study proposes the improved NPP estimating model (EXP-CASA) as follows:
\begin{equation}
\begin{aligned}
NPP & =  LUE_{max} \times FPAR \times Rad \times S_w(W) \times S_t(T) \times 0.5 \\ 
& =  LUE_{max} \times \frac{1}{\gamma_w} \times \frac{1}{\gamma_t}\times 0.5 \times VI^{\alpha_{v}} \times Rad \times{\alpha_{w}}^{W}\cdot W^{\beta_{w}} \times {\alpha_{t}}^{T}\cdot T^{\beta_{t}} \\ 
   & = \alpha_0 \times VI^{\alpha_{v}} \times Rad \times{\alpha_{w}}^{W}\cdot W^{\beta_{w}} \times {\alpha_{t}}^{T}\cdot T^{\beta_{t}}
\end{aligned}
\end{equation}
where $\alpha_0$ represents light use efficiency adjustment factor; W, T represent normalized LSWI and land surface temperature, respectively; $S_w(W)$ and $S_t(T)$ respectively represent $W_s$ and $T_s$ in Equation \ref{eq:CASA}; $\alpha_w,\beta_w$ represent moisture adjustment factor, which are used to show the photosynthetic inhibition of vegetation in very dry and overwatered environments; $\alpha_t,\beta_t$ represent temperature adjustment factor, which are used to show the photosynthetic inhibition of vegetation in low and high temperature environments.

The model involves new parameters, including a light use efficiency adjustment factor ($\alpha_0$), a vegetation index saturation effect limiting factor ($\alpha_v$), temperature adjustment factors ($\alpha_t,\beta_t$), and moisture adjustment factors ($\alpha_w,\beta_w$). To solve these parameters, we conducted the following analysis on the model in logarithmic space as Equation \ref{eq:lnNPP_2p}:
\begin{equation}
\label{eq:lnNPP_2p}
\begin{aligned}
    \ln NPP &= \ln( \alpha_{0} \cdot FPAR \cdot Rad \cdot S_w(W) \cdot S_t(T))\\
 &=\ln\alpha_0 + \alpha_{v} \ln VI + \ln Rad + (W\ln{\alpha_{w}} + \beta_{w}\ln W ) + (T\ln{\alpha_{t}} + \beta_{t}\ln T )\\
 & = \begin{bmatrix}
1, & \ln VI, & \ln Rad, & W, & \ln W, & T, & \ln T
\end{bmatrix}
\cdot 
\begin{bmatrix}
\ln \alpha_0 \\ 
\alpha_{v}\\ 
1\\ 
\ln \alpha_w \\ 
\beta_w\\ 
\ln \alpha_t\\ 
\beta_t
\end{bmatrix}
\end{aligned}
\end{equation}

By analyzing the EXP-CASA model in logarithmic space, we achieved a distinction between the model inputs and the parameters to be estimated. Specifically, the vegetation index, solar radiation, normalized LSWI, and temperature were designated as model inputs, whereas $\alpha_0,\alpha_v,\alpha_w,\beta_w,\alpha_t,\beta_t$ were identified as parameters for estimation. This approach effectively delineated the factors within the multiplicative model and facilitated the distinct handling of model inputs and estimable parameters. Consequently, this separation allows for the efficient use of observed NPP and multiple least squares regression to fit model parameters and calibrate the model.

\subsection{Evaluation of model performance metrics}
This research uses the following three accuracy indicators to evaluate the performance of EXP-CASA (and three NPP data products) in estimating NPP at flux observation sites, with the calculation equations as follows:
\begin{equation}
    \begin{aligned}
        R^2 &= \frac{\sum_{i=1}^{N}\left ( X_{NPP}-\overline{X_{OBS}} \right )^{2}}{\sum_{i=1}^{N}\left ( X_{OBS}-\overline{X_{OBS}} \right )^{2}} = 1 - \frac{\sum_{i=1}^{N}\left ( X_{NPP} - X_{OBS} \right )^{2}}{\sum_{i=1}^{N}\left ( X_{OBS}-\overline{X_{OBS}} \right )^{2}}\\
        RMSE &= \sqrt{\frac{\sum_{i=1}^{N}\left ( X_{NPP}-X_{OBS} \right )^{2}}{N}}\\
        Bias &= \frac{\sum_{i=1}^{N}\left ( X_{NPP}-X_{OBS} \right )}{N\times\overline{X_{OBS}}}
    \end{aligned}
\end{equation}
where $R^2$, $RMSE$ and $Bias$ are the R-squared, root-mean-square errors, and the difference between the estimated and observed values and the observed mean. The subscripts $X_{NPP}$ and $X_{OBS}$ denote the model's simulated and site's observed values, respectively.

\section{Results}
\label{Results}
In this study, given the heterogeneous spatiotemporal distribution of flux site quantities, we employed varied training samples at the site level to calibrate parameters, assess the EXP-CASA model's adaptability to diverse vegetation indices, and verify its stability and accuracy. Further analysis involved comparing the model with other products across various ecological categories, latitudinal zones, seasons, and chronological site data to evaluate the EXP-CASA model's effectiveness and accuracy.

\subsection{Model parameter fitting results and stability assessment}
\label{sec:random}
\subsubsection{Assessment of the adaptability of different vegetation indices to EXP-CASA model}

To analyze the impact of different vegetation indices on the EXP-CASA model and to identify and apply the most suitable one for the EXP-CASA model, this study compared three widely used vegetation indices in biomass estimation (NDVI, kNDVI, NIR$_v$, where kNDVI includes two variants as shown in Equation \ref{eq:kNDVI}). 

The comparative experiment was conducted 100 times, each time randomly selecting 70\% of the data for training and allocating the remaining data for testing. During these trials, multiple least squares regression fittings of parameters and NPP estimation were performed based on various vegetation indices of the proposed EXP-CASA model.

\begin{figure}[ht]
    \centering
    \includegraphics[width = \textwidth]{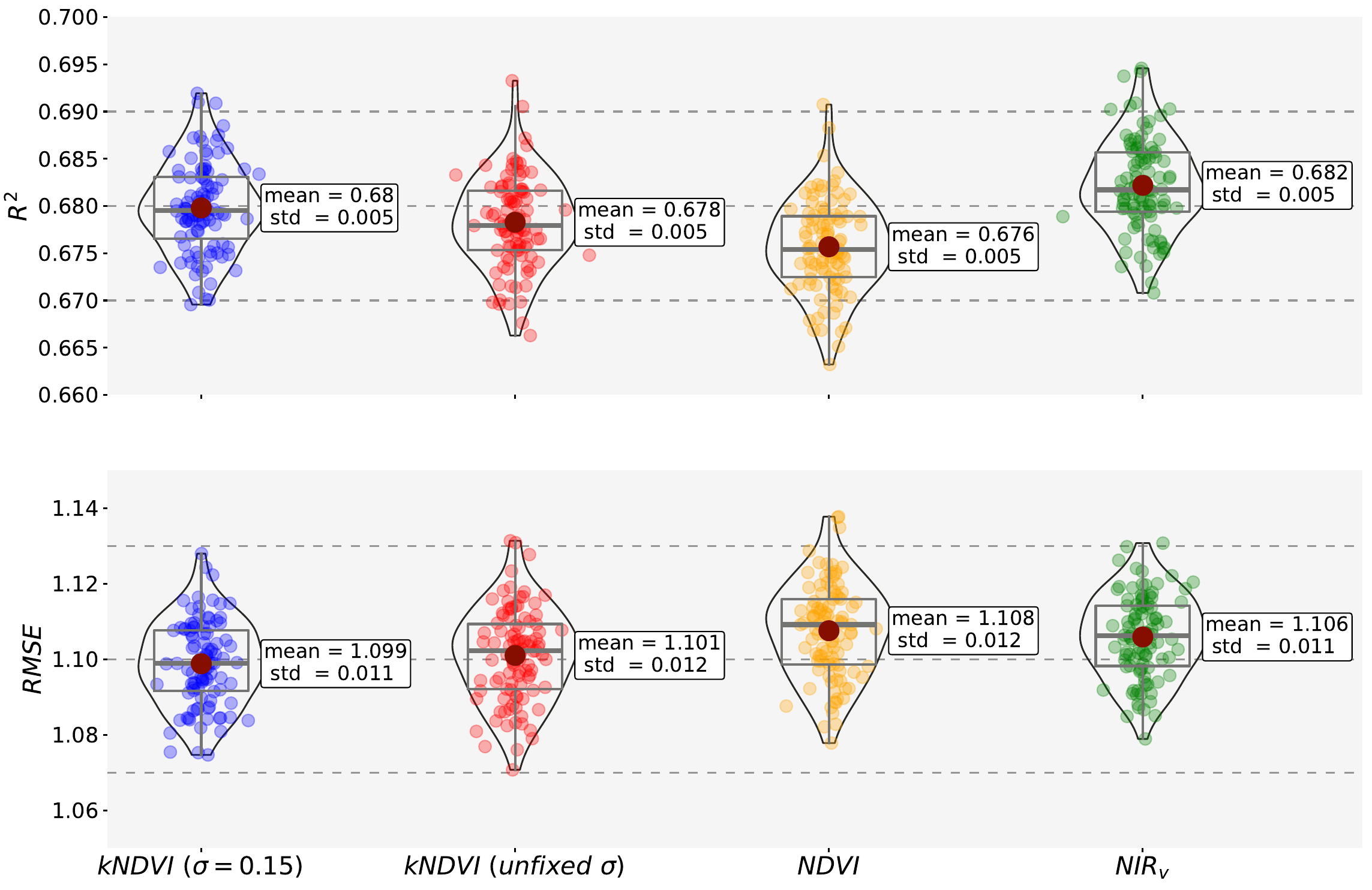}
    \caption{The statistical performance of the EXP-CASA model in estimating NPP with different vegetation indices in random experiment for 100 times, where unfixed $\sigma$ represtnts $\sigma = 0.5\cdot(\rho_{NIR} + \rho_{RED})$.
    }
    \label{fig:Results_rdmModelTest}
\end{figure}

Figure \ref{fig:Results_rdmModelTest} records the performance of the EXP-CASA model with different vegetation indices in estimating NPP on the test set. Results from random experiments indicate that the EXP-CASA models based on different vegetation indices consistently show good performance in terms of NPP estimation accuracy and stability. The mean and standard deviation of $R^2$ and $RMSE$ for all models remain at a consistent level, which indirectly reflects the potential correlation between vegetation indices and biomass. It is important to note that the non-linear vegetation index demonstrates superior compatibility with the EXP-CASA model over the NDVI. Further, the NPP estimation results from the EXP-CASA model, utilizing both kNDVI and NIR$_v$, show comparable performances. Considering the solid mathematical foundation and enhanced interpretability of kNDVI, it underscores the considerable promise kNDVI holds for integration into the EXP-CASA model. Accordingly, it is our reasoned recommendation to adopt kNDVI, anchored by a fixed $\sigma$ value, for the estimation of FPAR. To ensure unambiguous communication throughout this article, any further references to the EXP-CASA model will denote the  kNDVI, with the specified fixed $\sigma$ value 0.15.

\subsubsection{Evaluation of model temporal stability and parameter sensitivity}
Considering that the EXP-CASA model is data-driven, this study evaluates the temporal sensitivity of the model parameters. Based on the flux observation data from 2001-2014, we used the overlapping observation sequences of every two years as training samples and the remaining twelve years of data as testing samples, forming 13 sets of comparative experiments. At the same time, based on $\alpha_0$, stress control factors ($\alpha,\beta$), and normalized scaling coefficients $\gamma$, this study estimated the optimal moisture condition $W_{opt}$, the optimal temperature $T_{opt}$ for vegetation photosynthesis (which represent the normalized LSWI and temperature, respectively), and the potential LUE$_{max}$ for vegetation photosynthesis. By analyzing the parameter adaptation of different time series samples to other time periods, the temporal stability and sensitivity of the model parameters are evaluated.

\begin{table}[ht]
  \centering
  \caption{Table of experimental results on the temporal sensitivity of model parameters. $\mu,\sigma$ respectively represent the mean and standard deviation. }
    \begin{tabular}{ccccccccc}
    \toprule
        & $R^2$ & $RMSE$ & $Bias$ & $\alpha_{v}$ & $LUE_{max}$ & $W_{opt}$ & $T_{opt}$ \\
    \midrule
    mean ($\mu$)  & 0.67  & 1.12  & -0.08  & 0.36  & 0.60  & 0.73  & 0.54  \\
    std ($\sigma$)  & 0.01  & 0.02  & 0.03  & 0.10  & 0.07  & 0.05  & 0.03  \\
    min   & 0.64  & 1.10  & -0.13  & 0.15  & 0.50  & 0.67  & 0.49  \\
    max   & 0.68  & 1.16  & -0.03  & 0.55  & 0.70  & 0.82  & 0.58  \\
    $\mu - 2\sigma$ & 0.65 & 1.08 & -0.14 & 0.16 & 0.46 & 0.63 & 0.48 \\
    $\mu + 2\sigma$ & 0.69 & 1.16 & -0.02 & 0.56 & 0.74 & 0.83 & 0.60 \\
    \bottomrule
    \end{tabular}
  \label{tab:Results_Sensitivity}
\end{table}

Table \ref{tab:Results_Sensitivity} presents the results of the temporal sensitivity tests, showcasing the EXP-CASA model's accuracy metrics for the test dataset from corresponding experiments. With an $R^2 = 0.67 \pm 0.02, RMSE = 1.12 \pm 0.04 gC\cdot m^{-2} \cdot d^{-1}, Bias = -0.08 \pm 0.03$, the model demonstrates strong performance, highlighting its ability to produce consistent results across varied temporal datasets and indicating its robustness to time series variations. Furthermore, the vegetation index saturation effect adjustment factor, $\alpha_v = 0.36 \pm 0.2$, reveals moderate stability, suggesting varied vegetation saturation effects across time, yet within expected boundaries. This study also successfully determined the optimal moisture ($W_{opt} = 0.73 \pm 0.1$, corresponding LSWI is 0.46$\pm$0.2) and temperature ($T_{opt} = 0.54 \pm 0.06$, corresponding temperature is 288.04$\pm$3.9$K$) for vegetation photosynthesis under stress, at normalized conditions. These findings point towards a consistent optimal photosynthetic environment for vegetation, validating the stress estimation functions formulated in this research.

\begin{figure}
    \centering
    \includegraphics[width=\textwidth]{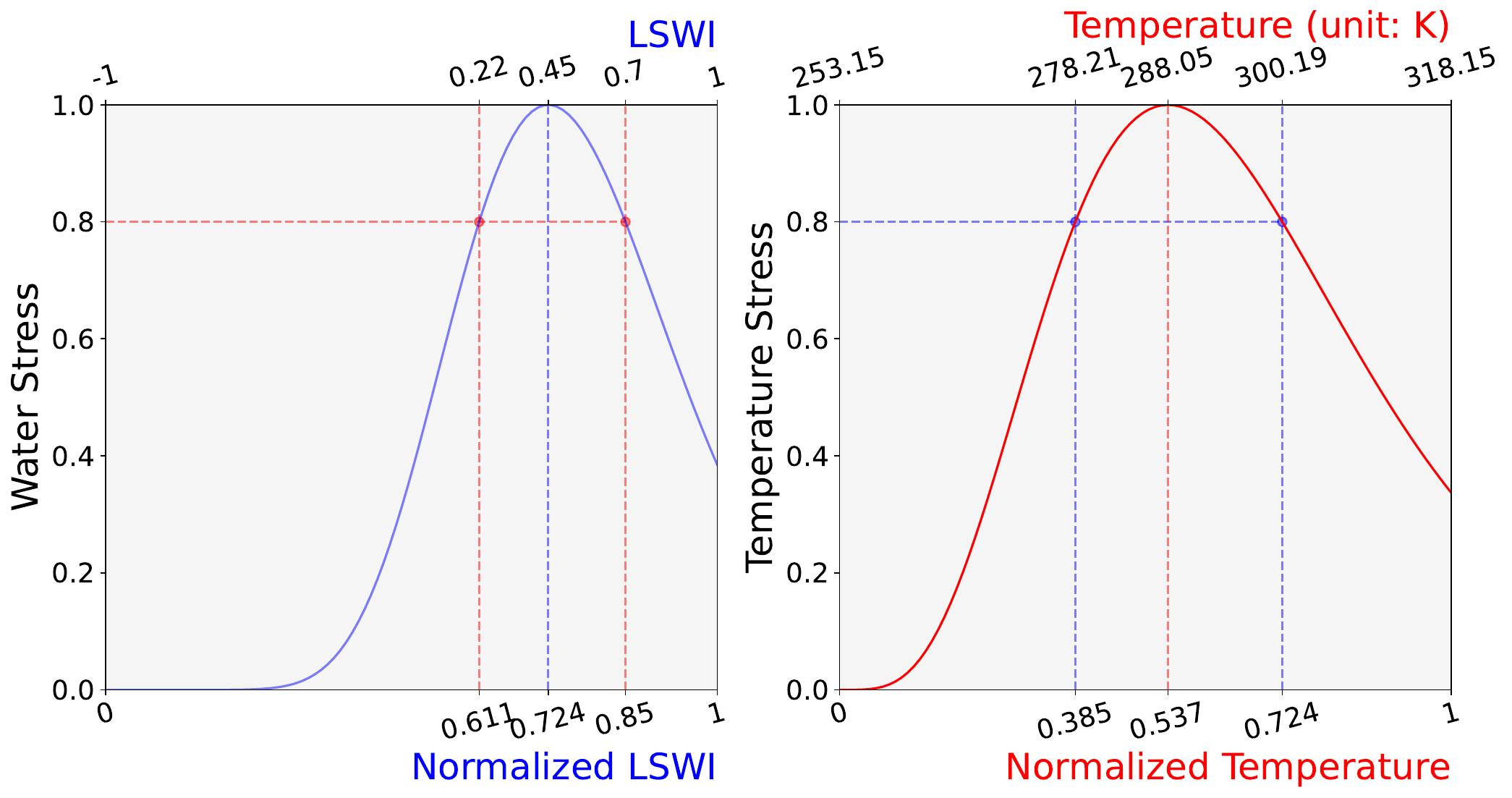}
    \caption{Temperature and water stress curves depicted by the EXP-CASA model, with optimal conditions for vegetation growth clearly marked, as recommended in this study, incorporate the critical moisture and temperature interval that is potentially ideal for vegetation photosynthesis.}
    \label{fig:curveStress}
\end{figure}

Overall, the EXP-CASA model constructed based on kNDVI demonstrates strong robustness and stability over time, with its parameters having low sensitivity to time series. Under different time series conditions, the optimal temperature and moisture conditions for vegetation photosynthesis inferred by the model are consistent, indicating the temporal consistency of the EXP-CASA model in estimating stress. At the same time, the potential LUE$_{max}$ obtained from model estimation is consistent with other studies (detailed in Section \ref{sec:lue_uncertainty}), showing the feasibility of the EXP-CASA model. This study fits a set of model parameters based on all observed data and proposes a complete innovative model for application and promotion as follows:

\begin{equation}
\label{eq:finalModel}
    NPP = exp (27.761-22.624\cdot W-8.423\cdot T)\times kNDVI^{0.381} \times W^{16.375} \times T^{4.523}\times Rad 
\end{equation}
where W, T represent normalized LSWI and surface temperature, respectively, and Rad represents solar radiation. Moreover, we derived the moisture and temperature stress curves which can be seen in Figure \ref{fig:curveStress} using Equation \ref{stress}. It is determined that in a moisture environment characterized by an LSWI range of 0.22 to 0.7 and a temperature range of 278.21K to 300.19K with the stress values of both greater than 0.8, vegetation experiences minimal stress effects and exhibits enhanced photosynthetic efficiency. Meanwhile, analysis reveals that the optimal water and temperature conditions, as indicated by comprehensive data, align closely with the values from Table \ref{tab:Results_Sensitivity}. This consistency across the experimental time series suggests that the stress curves are stable, highlighting their applicability in future studies.

\subsection{Compared to other products}
\label{sec:compare}

To further evaluate the accuracy and reliability of the NPP estimated by the EXP-CASA model, this study engaged in a comparative evaluation of the site-specific NPP estimation capabilities of EXP-CASA (outlined in Equation \ref{eq:finalModel}) against those offered by existing products such as MOD17A2H-NPP (hereafter referred to as MODIS-NPP) and GLASS-NPP, alongside the estimations derived from CASA model (hereafter referred to as CASA). Through this comparison with publicly available data products, the study validated the EXP-CASA model's performance and delved into its applicability and inherent limitations under varying environmental conditions. This study conducted a comparative analysis of the NPP estimation performance between the EXP-CASA model and other products or simulation data across four dimensions: varying ecological categories, diverse latitudinal ranges, distinct seasons, and temporal sequences at different sites.

\subsubsection{Comparison across different ecological categories}
To assess the adaptability of the EXP-CASA model to different ecosystems and effectively identify the model's strengths and weaknesses in various ecosystems, this study compared and analyzed the accuracy of NPP estimation by each model in different ecosystems. This research concentrates on six ecological categories (NF, EBF, DBF, MF, GRA, CRO) \footnote{The relatively low number of effective observations pertaining to the WET and SHR categories results in their exclusion from explicit presentation in the following comparative experiment.}. The classification system used originates from the International Geosphere-Biosphere Programme (IGBP), with detailed descriptions of these categories available in Appendix \ref{appendix_A}.
\begin{figure}[htbp]
    \centering
    \includegraphics[width = 0.85\textwidth]{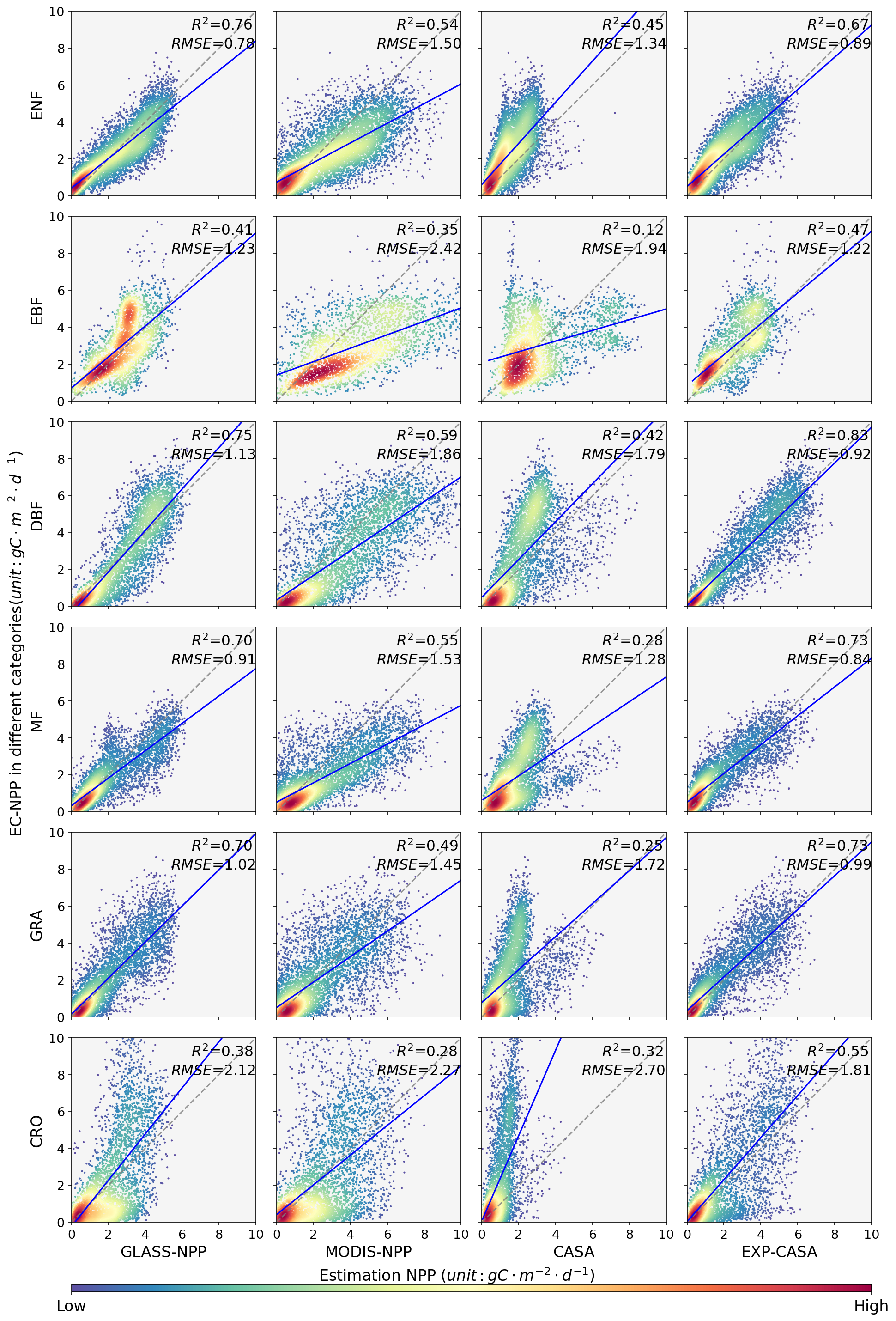}
    \caption{Scatter distribution and accuracy results of NPP estimated by different models under six major categories, where NF = Needle-leaf Forest, EBF = Evergreen Broadleaf Forest, DBF = Decidious Broadleaf Forest, MF = Mixed forest, GRA = Herbaceous, WET = Permanent Wetlands. The horizontal and vertical axes respectively represent EC-NPP and model-estimated NPP.}
    \label{fig:Results_Category}
\end{figure}

Figure \ref{fig:Results_Category} records the scatterplot distribution and accuracy results of the NPP estimated by each model under different ecological categories, where each row represents a different category and each column represents a different model; the horizontal and vertical axes represent EC-NPP and model-estimated NPP (unit: $gC\cdot m^{-2}\cdot d^{-1}$), respectively. 
The results show that although the precision indicators of EXP-CASA are slightly inferior to GLASS-NPP in needle-leaf forest (NF) category, the $R^2$ and $RMSE$ of EXP-CASA both perform the best in other categories. 
However, The NPP estimation performance of both MODIS-NPP and CASA across various categories is generally poor: MODIS-NPP's $R^2<0.6, RMSE_{max} = 2.42 gC\cdot m^{-2} \cdot d^{-1}$, and CASA's $R^2<0.5, RMSE_{max} = 2.7 gC\cdot m^{-2} \cdot d^{-1}$. 
Notably, in the evergreen broadleaf forest (EBF) category, both GLASS-NPP and EXP-CASA show poor estimation performance: lower $R^2$ while higher $RMSE$, together with significant underestimation, which might be caused by the vegetation index saturation effect due to the high canopy density of the evergreen broadleaf forests (EBF). The scatterplot distribution reveals that both EXP-CASA and GLASS-NPP accurately and consistently estimate NPP across various categories. However, within the cropland (CRO) and evergreen broadleaf forest (EBF) categories, which universally exhibit poor accuracy, EXP-CASA's estimations display a notably stronger correlation. In general, EXP-CASA has the capability to carry out consistent NPP estimations across different ecological categories.

\subsubsection{Comparison across different latitudes}
To assess the consistency of the EXP-CASA model at different latitudes and to quantify the model’s sensitivity to latitudinal zone differences, this study compared and analyzed the accuracy performance of various models in estimating NPP across different latitudinal distributions. Considering that the sites are distributed between 50\textdegree S and 70\textdegree N, this study divided the entire research interval into six latitude bands, with each band being 20\textdegree  wide, and compared the NPP estimating performance of four products across different latitude bands for quantitative analysis.

\begin{figure}[htbp]
    \centering
    \includegraphics[width = 0.85\textwidth]{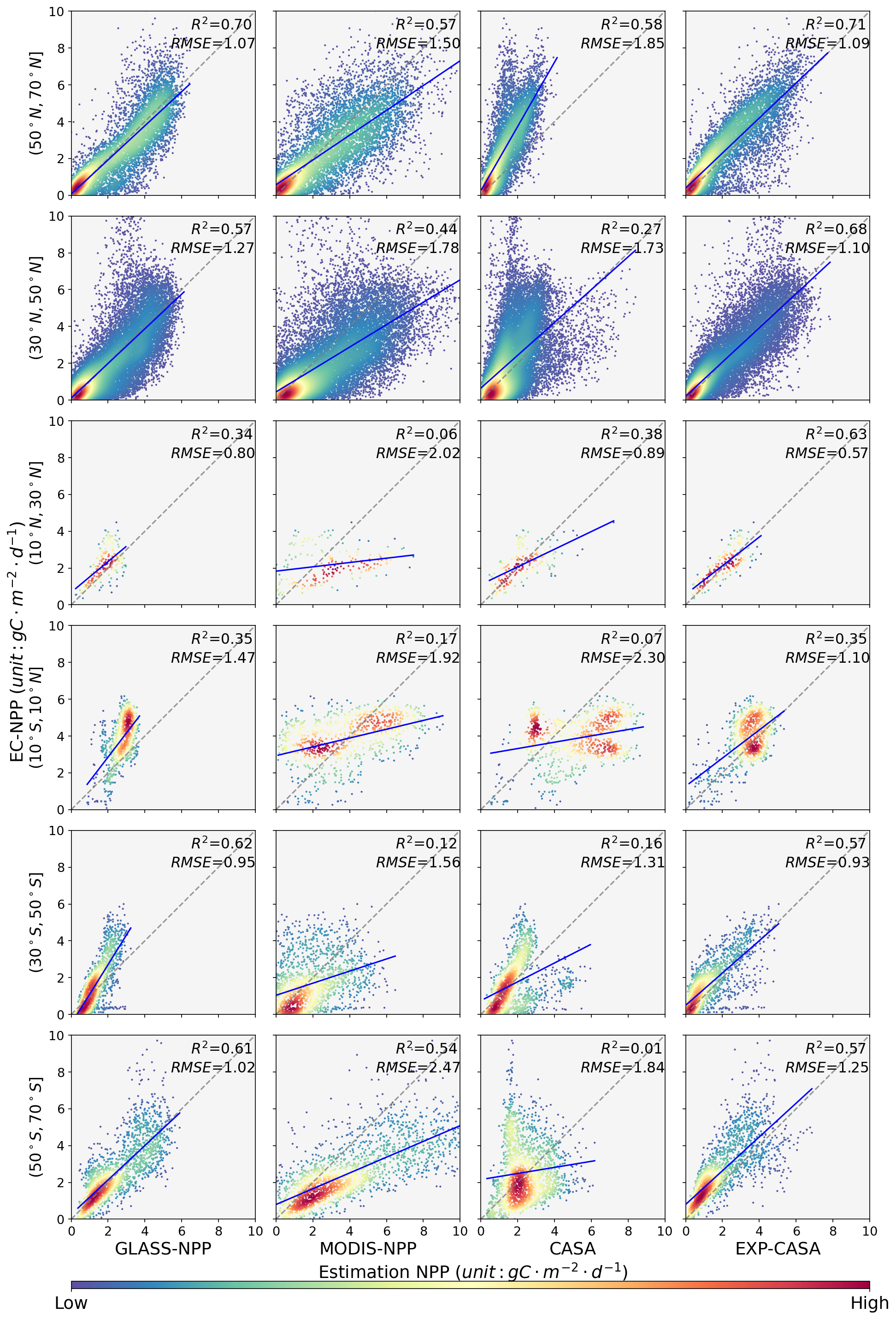}
    \caption{Scatter plots of NPP and EC-NPP from different models at various latitudinal regions, which are derived from the experimental bands divided every 20\textdegree  from 50\textdegree S to 70\textdegree N.}
    \label{fig:ResultsLatitude}
\end{figure}

Figure \ref{fig:ResultsLatitude} records the scatter distribution and accuracy results of NPP estimated by each model in different latitudinal regions, where each row represents a different latitudinal zone and each column represents a different model; the vertical axis and horizontal axis represent EC-NPP and the NPP estimated by the models, respectively (unit: $gC\cdot m^{-2}\cdot d^{-1}$). The results indicate that EXP-CASA shows the best performance in most latitudinal regions, only performs poor in the latitude zone from 30\textdegree S to 50\textdegree S; GLASS-NPP demonstrates relatively good estimation performance; whereas the estimation effects of MODIS-NPP and CASA are generally poor. It is noted that all models perform poor near the equator: the $R^2$ values of the four models are low while the $RMSE$ values are high. This may be due to the predominance of evergreen broadleaf forests (EBF) with high vegetation cover and biomass in the tropical rainforest areas near the equator, leading to higher saturation effect of vegetation index, which is also indicated by the significant underestimation by GLASS-NPP and EXP-CASA. From the scatter plots, EXP-CASA and GLASS-NPP are able to fit NPP more accurately in different latitude regions, having more ideal regression lines; in the latitude zone with the highest data abundance, from 30\textdegree N to 50\textdegree N, the estimating performance of EXP-CASA is far superior to the other three models. Overall, EXP-CASA exhibits good consistency across latitudes, effectively simulating the productivity changes brought by latitude zone differences.

\subsubsection{Comparison across different seasons}
To evaluate the consistency of the EXP-CASA model across different seasons and quantify the model's sensitivity to cyclical changes, this study compares and analyzes the accuracy of NPP estimations by various models across different seasons. According to the meteorological definition, the observational year at the site is divided into four seasons (with spring and autumn occurring at opposite times in the Northern and Southern Hemispheres). The study compares the NPP estimation performance of four products (showing in different column) in different seasons (showing in different row) and conducts a quantitative analysis.

\begin{figure}[htbp]
    \centering
    \includegraphics[width = 0.85\textwidth]{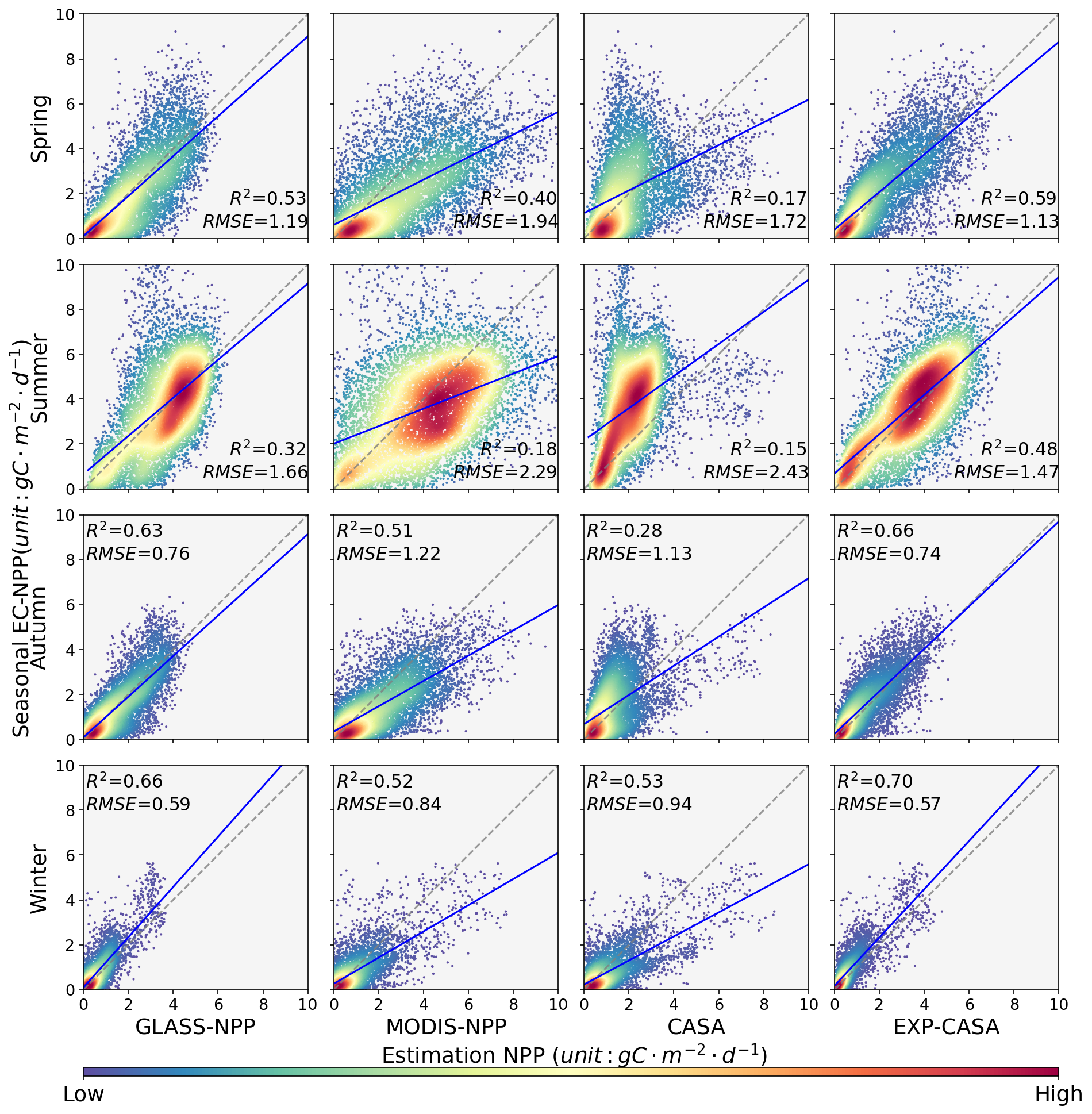}
    \caption{Scatter plots of NPP and EC-NPP from different models across different seasons, with the definition of seasons treated according to meteorology (March-May as spring, and so on). At the same time, the seasonal differences between the Northern and Southern Hemispheres are also taken into consideration.}
    \label{fig:Results_Season}
\end{figure}

Figure \ref{fig:Results_Season} records the scatter distribution and accuracy results of NPP estimated by various models in different seasons, where each row represents a different season and each column represents a different model; the vertical and horizontal axes represent EC-NPP and the NPP estimated by the models, respectively (unit: $gC\cdot m^{-2}\cdot d^{-1}$). The results show that EXP-CASA has the best accuracy performance, with $R^2$ and $RMSE$ superior to other models in every season. Meanwhile, MODIS-NPP and CASA fail to accurately concentrate on the simulation of NPP in all four seasons: both have $R^2$ values less than 0.55 and $RMSE_{max}$ exceeding 2.2 $gC\cdot m^{-2} \cdot d^{-1}$, significantly weaker than EXP-CASA and GLASS-NPP. It is noted that EXP-CASA and GLASS-NPP perform a consistent underestimation in winter, which might be due to the underestimation of vegetation indices caused by weather reasons such as ice and snow affecting some sites during winter. From the distribution of scatter plots, all models show a more dispersed distribution in summer, which may also be due to the vegetation index saturation effect caused by increased vegetation density during summer. At the same time, in the summer season where all models display dispersion, EXP-CASA is also able to form a relatively concentrated and stable simulation. Overall, EXP-CASA can better reflect the productivity changes brought about by seasonal differences.

\subsubsection{Comparison across different site time series}
NPP observations at sites often fluctuate, and a well-performing NPP estimation model should be able to capture these variations and reflect them in the fitting results. To assess the sensitivity of the EXP-CASA model to actual NPP changes within a specific spatial and temporal range, this study compared and analyzed the accuracy performance of different models in estimating NPP at typical flux sites. Furthermore, this study compared the sensitivity of GLASS-NPP and EXP-CASA to temporal NPP fluctuations at typical flux sites. Typical flux sites are derived from sites within the main ecological categories of this study that the ratio of the total number of valid observations to the total number of theoretical observations at the site is greater than 0.6.

\begin{figure}[htbp]
    \centering
    \includegraphics[width = \textwidth]{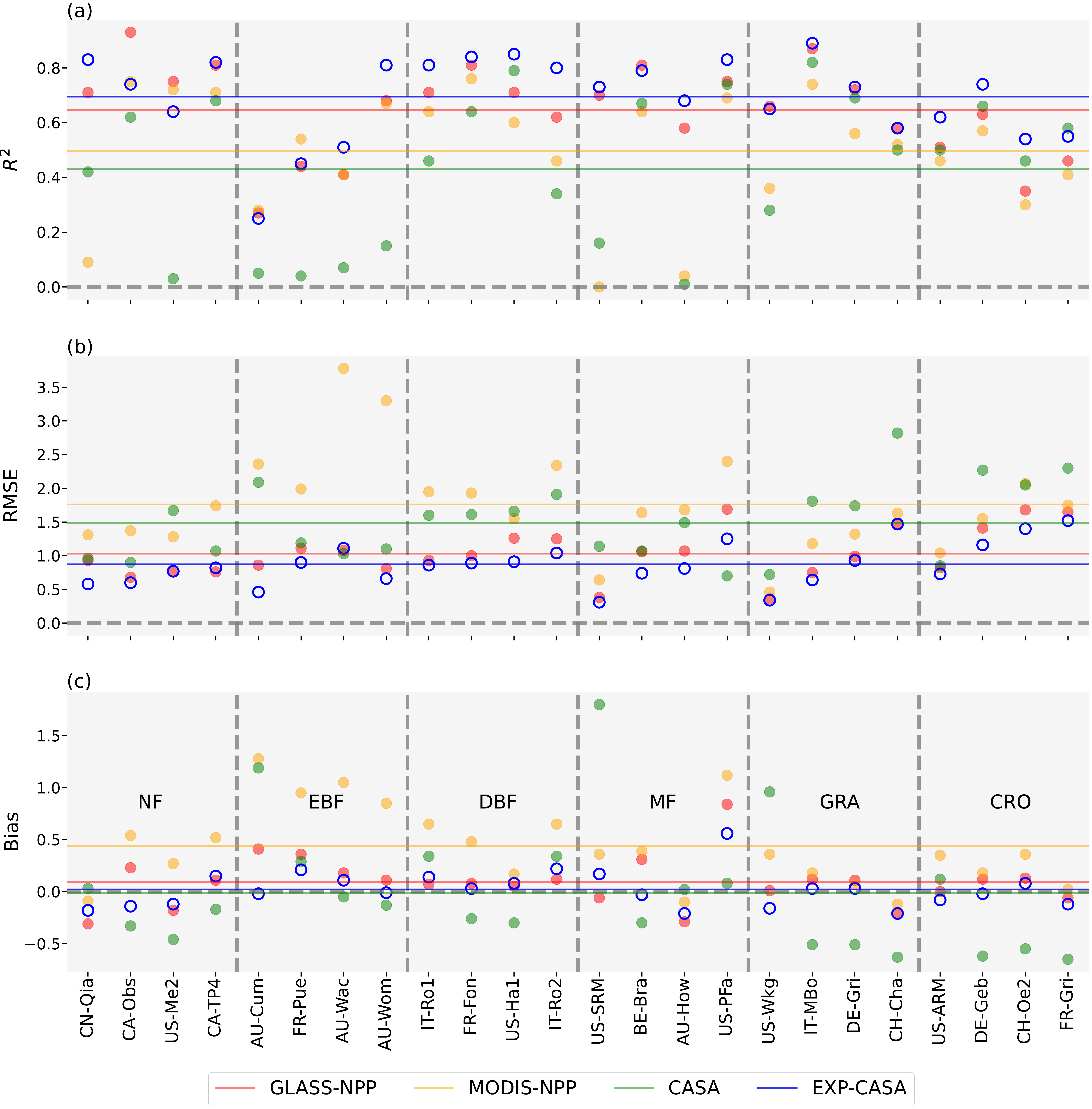}
    \caption{The distribution of statistical indicators at 24 typical observation sites across six major categories for the EXP-CASA and other NPP products. Sites with a significant number of effective observations and strong observational continuity were selected for analysis.}
    \label{fig:Results_SeriesStatistics}
\end{figure}

The results in Figure \ref{fig:Results_SeriesStatistics} document the accuracy distribution of various models across 24 typical observation sites, indicating that EXP-CASA outperforms other models in accuracy at the majority of typical observation sites, maintaining lower levels of $RMSE$ and $Bias$, with $R^2$ also taking the lead at most sites. It is noted that, although EXP-CASA and GLASS-NPP keep $RMSE$ and $Bias$ at lower levels across all observation sites, their $R^2$ levels are not high at some sites in evergreen broadleaf forests (EBF) and cropland (CRO), which is consistent with the category differences shown in Figure \ref{fig:Results_Category}.

\begin{figure}[htbp]
    \centering
    \includegraphics[width = \textwidth]{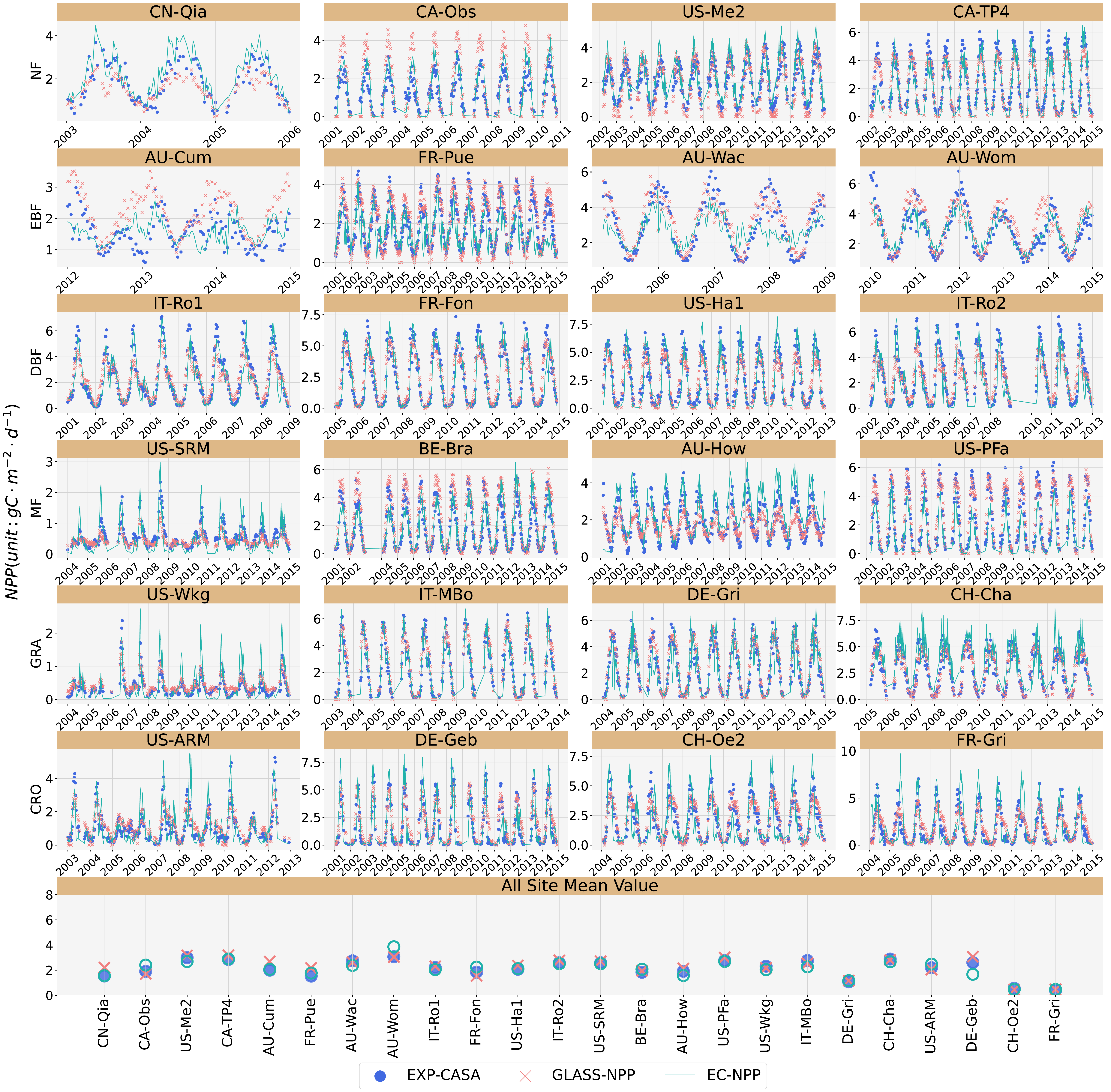}
    \caption{The temporal variation of the 8-day average NPP observed by EXP-CASA, GLASS-NPP, and 24 typical flux sites in time series. "All Sites" represents the average NPP corresponding to each site.}
    \label{fig:Results_Series}
\end{figure}

Figure \ref{fig:Results_Series} records the temporal changes of NPP at typical flux sites for EXP-CASA, GLASS-NPP, and EC-NPP, where the horizontal and vertical axes represent the year and NPP (unit: $gC\cdot m^{-2}\cdot d^{-1}$) respectively. The results indicate that EXP-CASA can accurately respond to NPP disturbances at the vast majority of experimental sites. Compared to GLASS-NPP, it is more consistent in estimating the actual observed NPP peak and trough values. Compared to other categories, evergreen broadleaf forest (EBF) sites experience very frequent disturbances in observed NPP values. Taking the AU-Cum site as an example: there exists an overall consistent periodic change trend, but short-term fluctuations are very large, with these larger variations occurring more frequently during periods of high biomass. In general, EXP-CASA maintains high accuracy in modeling the trends of periodical NPP changes and has high sensitivity to most short-term NPP disturbances. 

\section{Discussions}
\label{Discussions}
\subsection{Impact of saturation effects of vegetation indices on EXP-CASA}

\begin{figure}[ht]
    \centering
    \includegraphics[width = \textwidth]{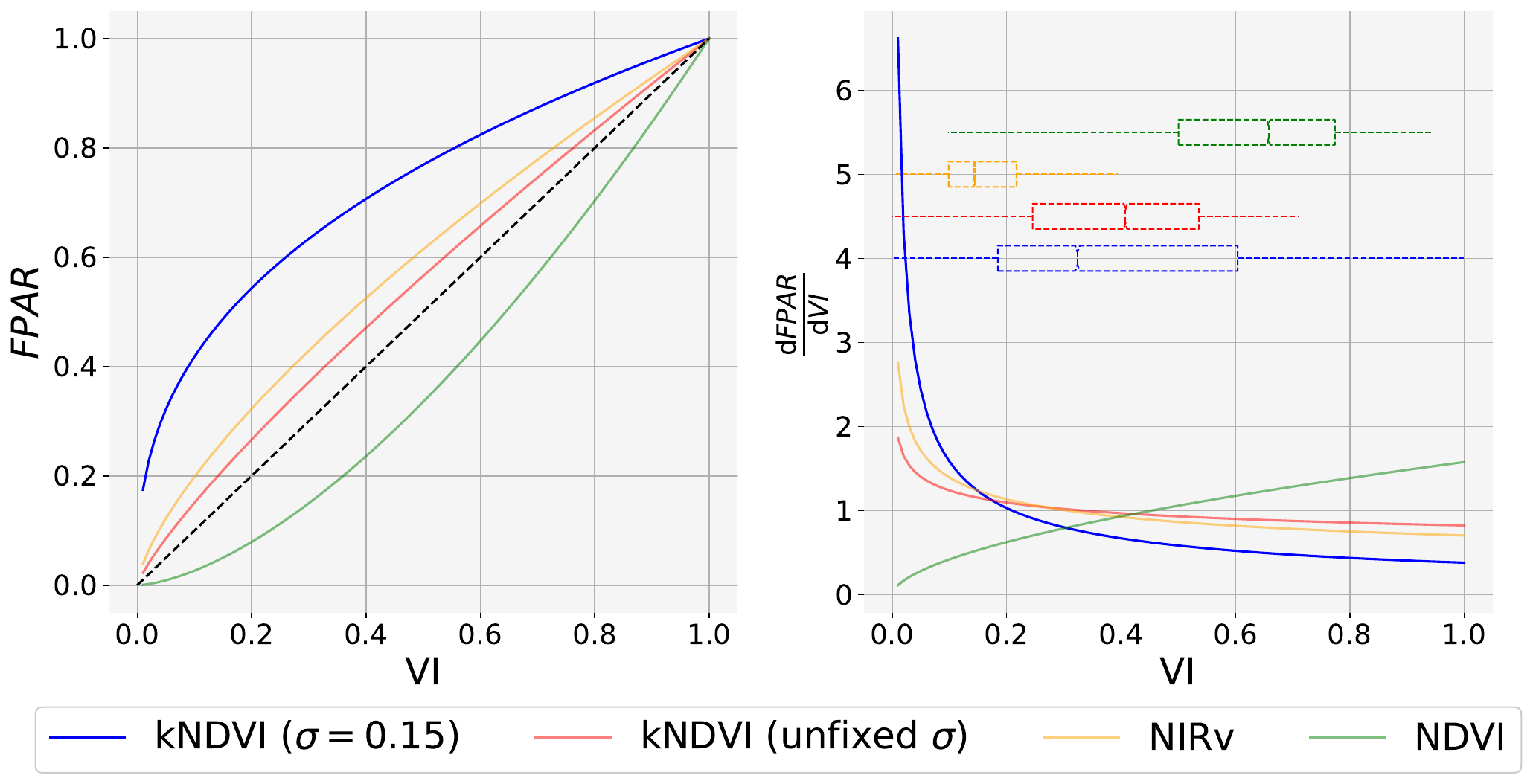}
    \caption{The curves of FPAR response to different vegetation indices and their gradient curves under the EXP-CASA model, where the horizontal axis represents the change in vegetation indices, and the vertical axes of the left and right graphs respectively represent FPAR and the gradient of FPAR change. At the same time, the right graph includes the numerical distribution of different vegetation index.}
    \label{fig:Discussion_fpar}
\end{figure}

Using the vegetation index as a direct proxy for FPAR is initially affected by the saturation effect of the vegetation index \citep{tian2020progress}. To mitigate this impact, we propose a vegetation index saturation correction factor $\alpha_v$, adjusting the saturation effect of vegetation index in a power form and quantifying its potential saturation situation. This form of estimating FPAR, similar to the direct ground vegetation NPP estimation based on the vegetation index, adequately considers the potential relationship between the vegetation index and NPP \citep{paruelo1997anpp,paruelo2000estimation}. Moreover, considering the properties of power functions, power exponents greater than 1 and less than 1 produce different curve morphologies. Figure \ref{fig:Discussion_fpar}, building on the random experiments in Section \ref{sec:random}, further visualizes the FPAR and FPAR gradient curves corresponding to different vegetation indices. The NDVI's saturation correction factor is greater than 1, indicating that compared to estimating FPAR through a linear model, the adjusted estimation model shows more significant variation at high NDVI values. This suggests that NDVI produces a higher saturation effect when vegetation canopy density is substantial, requiring a higher gradient of change at high NDVI values. For kNDVI $\sigma = 0.15$, kNDVI (with unfixed $\sigma$), and NIR$_v$, their saturation correction factors are all less than 1, suggesting that these three might exhibit excessive suppression of the saturation effect, requiring smaller gradients of change at high canopy densities. Notably, the adjustment factors for NIR$_v$ and kNDVI (with unfixed $\sigma$) are close to 1, showing a high linear correlation in their FPAR variation curves, demonstrating good resistance to saturation \citep{badgley2017canopy,zeng2022optical,gao2023evaluating,wang2024progress}.

The results of Section \ref{sec:compare} demonstrate the impact of VI saturation effects on NPP estimation. Within the spatial and temporal ranges of high canopy density, such as in summer, near the equator, and evergreen broadleaf forest (EBF), the CASA model, which directly uses NDVI as a proxy for FPAR, exhibits significant underestimation of NPP \citep{braghiere2020characterization,wang2024progress}. Although EXP-CASA adjusts for FPAR using a vegetation index saturation correction factor, its accuracy performance in areas with high canopy density is poor compared to regions of lower canopy density where the model generally excels, with more dispersed scatter distributions and lower sensitivity to high biomass fluctuations at sites. To delve deeper into the impact of $\alpha_v$ calibration on the saturation phenomena of vegetation indices, this study assessed the efficacy of EXP-CASA, utilizing both NDVI and kNDVI, in NPP estimation under conditions with and without $\alpha_v$ calibration. This comparison maintained a consistent stress estimation method across the same vegetation indices. As shown in Table \ref{tab:alphavCalibration}, the results reveal a marked improvement in the precision of overall estimation after calibrating by the Equation \ref{eq:FPAR}. This underscores the model's success in adjusting for the saturation effects inherent in NDVI, and its efficacy in addressing the potential over-suppression of saturation effects in kNDVI. In summary, the FPAR estimation method employed in this study has mitigated the saturation effect to a notable degree and successfully facilitated a quantitative evaluation of this phenomenon.

\begin{table}[htbp]
  \centering
  \caption{Comparative Performance of EXP-CASA Based on NDVI and kNDVI with and without $\alpha_v$ Calibration (according to Equation \ref{eq:FPAR}).}
    \begin{tabular}{cccc}
    \toprule
    Model & $\alpha_v$-Calibrated & $R^2$    & $RMSE$ \\
    \midrule
    EXP-CASA$_{kNDVI}$ & \Checkmark  & 0.680  & 1.100  \\
    EXP-CASA$_{kNDVI}$ & \XSolid  & 0.643  & 1.395  \\
    EXP-CASA$_{NDVI}$ & \Checkmark  & 0.676  & 1.108  \\
    EXP-CASA$_{NDVI}$ & \XSolid  & 0.673  & 1.164  \\
    \bottomrule
    \end{tabular}%
  \label{tab:alphavCalibration}%
\end{table}%

\subsection{Uncertainty in the estimation of the actual light use efficiency}
\label{sec:lue_uncertainty}

Table \ref{tab:Results_Sensitivity} records the category-independent potential maximum light use efficiency ($0.6 \pm 0.14 gC\cdot MJ^{-1}$) estimated by EXP-CASA based on time-series sensitivity experiments, which is theoretically consistent with the results of existing studies. In previous studies, the LUE$_{max}$ was considered a parameter differentiated by ecological category. Initial research on CASA suggested that the LUE$_{max}$ of global vegetation is a globally consistent parameter \citep{potter1993terrestrial,field1995global}, and the upper limit of vegetation maximum LUE is 3.5 $gC\cdot MJ^{-1}$ \citep{raymond1994relationship,zhu2006simulation}. Studies focusing on the differentiation of light use efficiency by vegetation category argue that the maximum LUE is a category-differentiated constant, with different categories of vegetation having maximum LUE values ranging between 0.389 and 1.259 $gC\cdot MJ^{-1}$ \citep{running2000global,zhu2006simulation,liu2019global,xiao2022estimation}. At the same time, the category-independent LUE$_{max}$ of EXP-CASA is not sensitive to time series, further indicating the accuracy and stability of the EXP-CASA model. Employing a category-independent LUE$_{max}$ approach might yield lower accuracy for specific categories when compared to using parameters that differentiate between categories. However, given the complexity and data requirements necessary to estimate and calibrate category-differentiated parameters, this reduction in accuracy is deemed acceptable.

At the same time, the optimal photosynthetic moisture and temperature conditions estimated by EXP-CASA, which are category-independent, remain stable over time. However, given the significant differences in optimal growth conditions among different vegetation types, the potential variation in conditions for the same vegetation type across different spaces, and the uneven distribution of flux observation data across categories, these category-independent conditions may introduce potential global errors. This study further explored a simplified form of the model under globally consistent optimal photosynthetic moisture and temperature conditions and analyzed the performance under category differentiation (detailed in Appendix \ref{appendix_B}). The category-differentiated EXP-CASA model does not significantly enhance estimation performance but considerably increases the model's complexity. In regions where vegetation categories are ambiguous, for example, when the used land cover categories do not match the model's differentiated categories, there exists a potential fixed $Bias$ in estimated NPP. Therefore, employing the globally consistent EXP-CASA model represents the globally optimal solution under current data and experimental conditions.

\subsection{Limitations of the EXP-CASA model}
The EXP-CASA model exhibits commendable performance at the site level, nonetheless, it has inherent limitations. Parameter fitting and model calibration heavily depend on flux observation site data. Nonetheless, the limited temporal span and geographically sparse coverage of open-source flux observation data severely challenge the model's calibration capabilities and the assessment of product quality. Despite EXP-CASA achieving a high $R^2$ and low $RMSE$ in alignment with EC-NPP across various experimental settings, it consistently underestimates ($Bias$ $<$ 0), suggesting potential systematic $Bias$ within the EXP-CASA framework. However, the introduction of a fixed correction coefficient to counteract this $Bias$ could potentially alter the model’s fundamental linear correlation with EC-NPP in logarithmic space. Furthermore, the EXP-CASA model calibrates FPAR to account for the vegetation index saturation effect employing a power function. However, this method universally applied calibration risks obscuring the potential linear relationship between FPAR and vegetation index in areas with low canopy density, though this deficiency has yet to be conclusively proven. Despite these challenges, our model maintains a straightforward form and exhibits robust performance across diverse evaluation limitations. 

\section{Conclusions}
\label{Conclusions}
The unavoidable saturation effects of the vegetation index have posed a significant limitation on the estimation of NPP using the classic CASA model. The classic CASA model requires complex, multi-source data and parameter inputs, which introduces more uncertainty into NPP estimation. Therefore, it is necessary to optimize current large-scale NPP estimation methods to reduce errors in primary productivity estimation. This study, based on the classic CASA model and flux observation data, proposes an improved NPP estimation model (EXP-CASA). It improves the estimation method of FPAR, simulating the non-linear relationship between the vegetation index and FPAR in the form of a power function, and evaluating the saturation effects of the vegetation index. At the same time, it improves the estimation method of environmental stress, reflecting the inhibitory effect of extreme environments on vegetation photosynthesis and the optimal photosynthesis efficiency under the most favorable conditions with a consistent change curve. Furthermore, this study assesses the model's adaptability to different vegetation indices and parameter sensitivity to time series, and compares the performance of four NPP products across different seasons, latitudinal zones, ecological categories, and site time series.

Our research shows that: The EXP-CASA model performs good adaptability to different vegetation indices, and its parameters are not sensitive to the time series. In the time-series stability analysis, the model estimated NPP with an $R^2$ value of $0.67 \pm 0.02$ and an RMSE of $1.12 \pm 0.04 \, gC \cdot m^{-2} \cdot d^{-1}$. At the same time, this study inversely estimated the potential maximum light use efficiency of vegetation ($LUE_{max} = 0.6 \pm 0.14$), the optimal moisture factor ($W_{opt} = 0.73 \pm 0.1$, corresponding LSWI is 0.46 $\pm$ 0.2), and the optimal
temperature factor ($T_{opt} = 0.54 \pm 0.06$, corresponding temperature is 288.04 $\pm$ 3.9$K$) for vegetation photosynthesis. These indicators all remained at a stable level, reflecting the stable performance of the model in estimating NPP. Secondly, in the comparison experiments of four different NPP products across different ecological categories, seasons, and latitudinal zone, EXP-CASA had the overall best NPP estimation performance, with its $R^2$ and $RMSE$ being superior to the other products under the majority of experimental conditions, followed by GLASS-NPP. On the other end of the spectrum, MODIS-NPP and the classic CASA were found to be less aligned with the observed performance indicators. Furthermore, EXP-CASA had the highest sensitivity to NPP anomalies at flux sites, capable of better simulating short-term minor fluctuations in NPP while reflecting periodical changes in NPP. This further reflects high stability and great performance of EXP-CASA in estimating site NPP.

In the future, we will extend EXP-CASA model to more accurate estimation scenarios. A pivotal area of focus will be on enhancing the handling of no-data values in reflectance data and thoroughly evaluating the effects of sensors, terrain, cloud cover, and other factors on the model's ability to estimate NPP over extensive spatio-temporal scales. It is noteworthy that, for more precise estimation scenarios, relying solely on the LSWI to estimate water stress proves to be insufficient. There's a pressing need to delve deeper into the impact of soil moisture and groundwater runoff on the water stress component of vegetation photosynthesis. Moreover, the flux observation data used for model parameter fitting are limited in time and unevenly distributed in space, making it difficult to make a more comprehensive and effective assessment of model performance. This requires obtaining more accurate, real-time flux observation data. In future research, we will focus on these issues, combining the model with reflectance data of Landsat and Sentinel-2 and utilizing thermal-based data to optimize temperature stress estimation, further analyzing the impact of terrain and scale on the estimation of vegetation net primary productivity.

\section*{Declaration of Competing Interest}

The authors declare that they have no known competing financial interests or personal relationships that could have appeared to influence the work reported in this paper.

\section*{Acknowledgments}

We are very grateful for the support of all the providers of open-source data. At the same time, we also appreciate the researchers who offered assistance during the process of writing the paper. Finally, we are also thankful for the reviewers' evaluation of our paper and the constructive comments they made.

\section*{Funding}
This work was supported by the National Natural Science Foundation of China [grant number 42101346]; the ``Unveiling and Commandin'' project in the Wuhan East Lake High-tech Development Zone [grant number 2023KJB212]; the China Postdoctoral Science Foundation [grant number 2020M680109]; and the Undergraduate Training Programs for Innovation and Entrepreneurship of Wuhan University (GeoAI Special Project) [grant number S202210486299].

\section*{Data availability}

Data will be made available on request.

\appendix
\section{NPP products and ecological categories overview}
\label{appendix_A}

Considering both the IGBP classification system and the vegetation categories included in FLUXNET observation sites, the vegetation ecological categories and their descriptions applied in this research are referred to in Table \ref{tableVegetationTypes}.

\begin{table}[ht]
  \centering
  \caption{Terrestrial vegetation ecological categories, abbreviations, and detailed descriptions used in the study.}
    \begin{tabular}{ccc}
    \toprule
    Vegetation Types & Acronym & Description \\
    \midrule
    Needleleaf Forest & NF    &  \begin{tabular}{p{8cm}}
    Dominated by conifer trees (canopy $>$2m) and tree cover $>$60\% 
    \end{tabular}\\
    Evergreen Broadleaf Forest & EBF  &  \begin{tabular}{p{8cm}} Dominated by evergreen broadleaf and palmate trees (canopy $>$2m). Tree cover $>$60\% \end{tabular}\\
    Deciduous Broadleaf Forest & DBF  &  \begin{tabular}{p{8cm}}Dominated by deciduous broadleaf trees (canopy $>$ 2m). Tree cover $>$60\% \end{tabular}\\
    Mixed Forests & MF    &  \begin{tabular}{p{8cm}}Dominated by neither deciduous nor evergreen tree type (canopy $>$2m). Tree cover $>$10\% \end{tabular} \\
    Shrublands & SHR   &  \begin{tabular}{p{8cm}} Dominated by woody perennials and tree cover$>$10\%\end{tabular} \\
    Grasslands & GRA   &  \begin{tabular}{p{8cm}}Dominated by herbaceous annuals ($<$2m)\end{tabular} \\
    Permanent Wetlands & WET   &  \begin{tabular}{p{8cm}}Permanently inundated lands with 30-60\% water cover and $>$10\% vegetated cover \end{tabular}\\
    Croplands & CRO   &  \begin{tabular}{p{8cm}}At least 60\% of area is cultivated cropland or small-scale cultivation 40-60\% with natural tree, shrub, or herbaceous vegetation \end{tabular} \\
    \bottomrule
    \end{tabular}%
  \label{tableVegetationTypes}%
\end{table}%

The GLASS-NPP product used in this research comes from the publicly available GLASS Net Primary Productivity product from the GLASS dataset\footnote{Global LAnd Surface Satellite (GLASS): \url{http://www.glass.umd.edu/}.}. The productivity calculation is carried out using an improved EC-LUE model \citep{yuan2007deriving,yuan2010global,yuan2014global}. The GLASS employs the latest version of the EC-LUE model (which has not been published yet), taking into account the effects of atmospheric $CO_2$ concentration, radiation components, and atmospheric vapor pressure deficit (VPD) on productivity. This can effectively reproduce spatial, seasonal, and annual variations, especially improving the performance of the model in replicating productivity changes. The GLASS-NPP is derived from the gross primary productivity of GLASS and the respiration index (the ratio of NPP to GPP), which is calculated from 19 dynamic global vegetation models (DGVMs).

The MODIS net photosynthesis product originates from MOD17A2H \citep{running2015modis,running2021modis}. This dataset is a 500-meter pixel size, 8-day cumulative Gross Primary Productivity (GPP) product based on the concept of radiation use efficiency. The dataset algorithm is derived from the MOD17 algorithm, which is key to expressing the limitations of water and temperature on actual light energy utilization rate in the form of piecewise functions. The net photosynthesis band value (PSN) is GPP minus the maintenance respiration (MR) \citep{running2015modis,running2021modis}. This product has been widely used in various studies and application analyses of vegetation productivity. However, the PSN data overlooked the respiratory consumption attributed to vegetation growth, and the MODIS-supplied actual NPP is presented annually. Consequently, in this study's comparative experiment, over an 8-day time scale, MODIS-NPP appears exaggerated, as evidenced by Section \ref{sec:compare}.

\section{Impact of category differentiation on the EXP-CASA model}
\label{appendix_B}

The results of this study indicate: the most suitable moisture and temperature conditions for vegetation photosynthesis remain stable over time. Furthermore, we propose a hypothesis that simplifies EXP-CASA further by fixing the optimum photosynthesis water and temperature factors for the vegetation. Let this optimum factor be denoted as c ($c = -\frac{\beta}{\ln \alpha}$), then the Equation \ref{eq:lnNPP_2p} can be transformed as follows:

\begin{equation}
\begin{aligned}
\ln NPP &= \ln( \alpha_{0} \cdot FPAR \cdot Rad \cdot W_s \cdot T_s)\\
 &=\ln\alpha_0 + \alpha_{v} \ln VI + \ln Rad + (W\ln{\alpha_{w}} + \beta_{w}\ln W ) + (T\ln{\alpha_{t}} + \beta_{t}\ln T )\\
 & = \ln \alpha_0 + \alpha_{v} \ln VI + \ln Rad + (W-c_w\ln W)\ln \alpha_w + (T-c_t\ln T)\ln \alpha_t\\
& =
\begin{bmatrix}
1, & \ln VI, & \ln Rad, & W-c_w\ln W, &T-c_t\ln T
\end{bmatrix}
\cdot 
\begin{bmatrix}
\ln \alpha_0 \\ 
\alpha_{v}\\ 
1\\ 
\ln \alpha_w \\ 
\ln \alpha_t\\ 
\end{bmatrix}
\end{aligned}
\label{eq:lnNPP_1p}
\end{equation}
where $c_w,c_t$ respectively represent the optimum moisture and temperature environments for vegetation photosynthesis, simplifying the model's parameter form further at this point. Meanwhile, assessing the application potential of category-differentiated models versus category-independent models is crucial to deciding on the optimum parameter model construction method.

In this study, we employed Equations \ref{eq:lnNPP_2p} and \ref{eq:lnNPP_1p} for the derivation of category-differentiated parameters and the estimation of NPP respective to the ecological categories of the observed sites. Crucially, the innovation of our research centers around the comparative experimentation process, wherein we evaluate the performance of NPP estimations across different ecological categories, employing both category-differentiated and category-independent EXP-CASA models for a comprehensive assessment. Experiments V1 and V3 exemplify the application of category-independent models, whereas V2 and V4 illustrate the utilization of category-differentiated models. The parameterization for experiments V1 and V2 was conducted in accordance with Equation \ref{eq:lnNPP_2p}, and for V3 and V4 following Equation \ref{eq:lnNPP_1p}. This structured approach enabled us to contrast the category-differentiated EXP-CASA model, devised through Equations \ref{eq:lnNPP_2p} and \ref{eq:lnNPP_1p}, against the category-independent EXP-CASA model in terms of NPP estimation efficiency across different ecological categories.

\begin{table}[htbp]
  \centering
  \caption{Overall NPP estimation performance of the EXP-CASA model under four experimental conditions, where Category independence represents whether the experimental model is independent of categories, while Model represents the formula of the experimental model.}
    \begin{tabular}{cccccc}
    \toprule
    Experiment & Category independence & Model & $R^2$    & $RMSE$  & $Bias$ \\
    \midrule
    V1    & \Checkmark  & Equation \ref{eq:lnNPP_2p}  & 0.68  & 1.10  & -0.08  \\
    V2    & \XSolid  & Equation \ref{eq:lnNPP_2p}  & 0.70  & 1.06  & -0.07  \\
    V3    & \Checkmark  & Equation \ref{eq:lnNPP_1p}  & 0.68  & 1.10  & -0.08  \\
    V4    & \XSolid  & Equation \ref{eq:lnNPP_1p}  & 0.68  & 1.09  & -0.07  \\
    \bottomrule
    \end{tabular}%
  \label{tab:SUP_CategoryExperimentsAll}%
\end{table}%

\begin{figure}[ht]
    \centering
    \includegraphics[width = \textwidth]{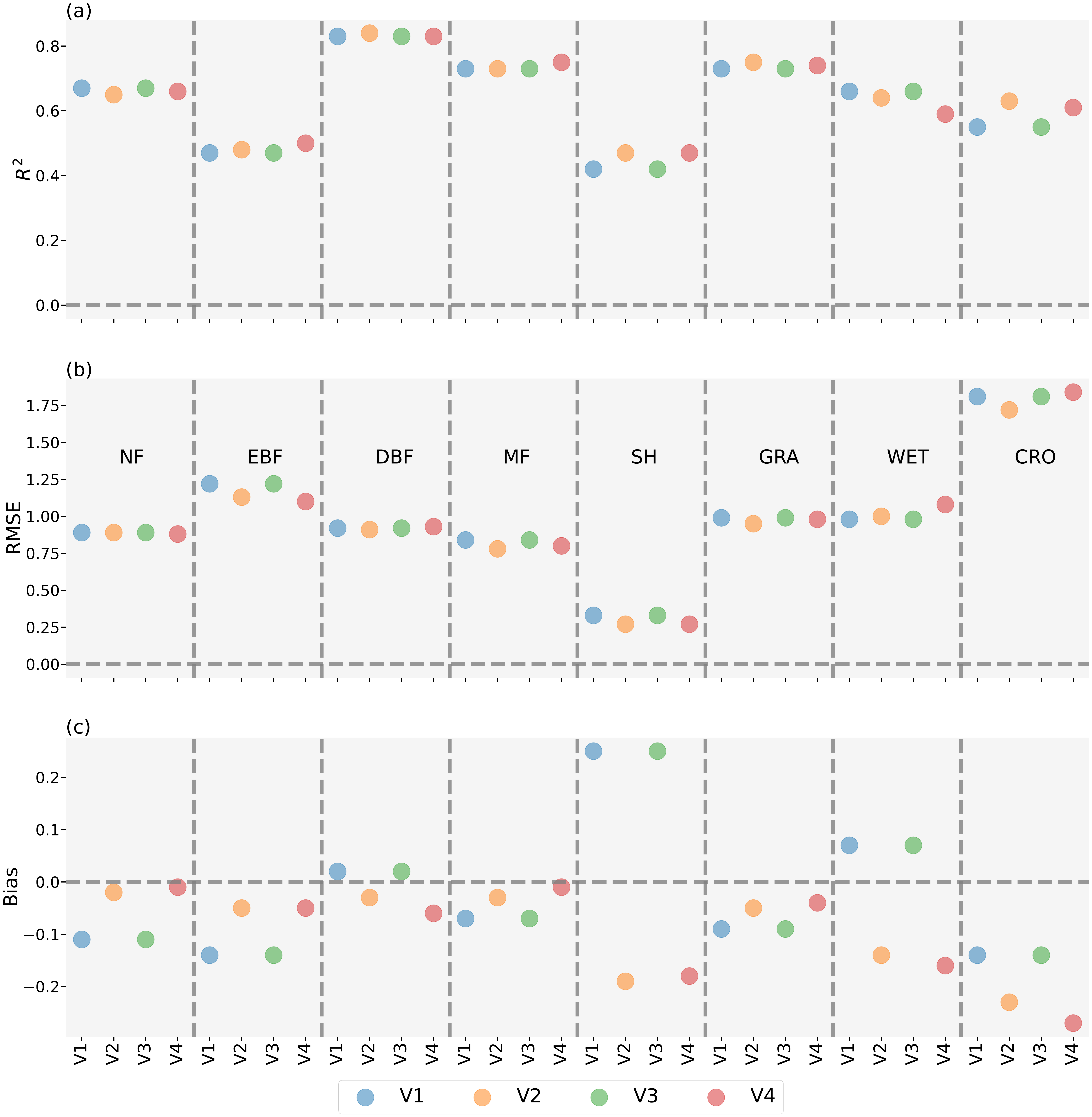}
    \caption{Performance of NPP estimation by the EXP-CASA model under different experimental conditions across different ecological categories, where V1 and V3 represent the category-independent models, V2 and V4 represent the the category-differentiated models, V1 and V2 are constructed by Equation \ref{eq:lnNPP_2p}, V3 and V4 are constructed by \ref{eq:lnNPP_1p}.}
    \label{fig:SUP_CategoryExperiments}
\end{figure}

Figure \ref{fig:SUP_CategoryExperiments} and Table \ref{tab:SUP_CategoryExperimentsAll} record the overall performance and the performance across different ecological categories of four variants from EXP-CASA in NPP estimation, respectively.
In terms of the performance evaluation metrics, the category-differentiated models (V2, V4) are slightly better than the category-independent models (V1, V3), but the difference in evaluation metrics is not significant. The four models perform similarly in terms of $R^2$, $RMSE$, and $Bias$. It is noted that all four models performed significant estimation errors in the cropland and evergreen broadleaf category \citep{wang2024progress}, suggesting that category differentiation is also difficult to further optimize the saturation effect of vegetation index brought about by estimating primary productivity based on canopy reflectance.

Overall, compared to category-independent models, category-differentiated models have led to an improvement in NPP estimation performance in some categories, but the extent of improvement is not significant. Considering that category differentiation requires more observed data to support model fitting and subsequent calibration, and it also introduces errors in identifying land cover categories, using the category-independent EXP-CASA is a reasonable global solutions.

\printcredits
\bibliographystyle{model1-num-names}
\bibliography{refs}
\end{document}